\documentclass[12pt,epsfig]{article}
\usepackage{epsfig}
\usepackage{amssymb}
\newcommand{\singlespace}{
     \renewcommand{\baselinestretch}{1}\large\normalsize}
\newcommand{\doublespace}{
     \renewcommand{\baselinestretch}{1.6}\large\normalsize}

\newcommand{\beq}{\begin{equation}}
\newcommand{\eeq}{\end{equation}}
\newcommand{\bea}{\begin{eqnarray}}
\newcommand{\eea}{\end{eqnarray}}

\newcommand{\ave}[1]{\langle {#1} \rangle}

\newcommand{\pslash}{p\!\!\!/}

\newcommand{\pb}{\bar\psi}
\newcommand{\pd}{\psi^{\dagger}}
\newcommand{\qq}{\ave{\pb\psi}}
\newcommand{\dg}{\delta g_{\pi qq}^{-2}}
\newcommand{\fig}[1]{Fig.~\ref{#1}}
\newcommand{\Sec}[1]{Sec.~\ref{#1}}
\newcommand{\eq}[1]{Eq.~(\ref{#1})}
\newcommand{\eqs}[1]{Eqs.~(\ref{#1})}
\newcommand{\nce}{$1/N_c$-expansion scheme}

\def\roughly#1{\mathrel{\raise.3ex\hbox{$#1$\kern-.75em%
\lower1ex\hbox{$\sim$}}}}
\def\lsim{\roughly<}
\def\gsim{\roughly>}
\def\psl{p\hspace{-1.7mm}/}

\def\dfp{\frac{d^4 p}{(2\pi)^4}}
\def\dfk{\frac{d^4 k}{(2\pi)^4}}
\def\intp{\int\dfp}
\def\intk{\int\dfk}
\def\rrr{\longrightarrow}
\def\={\;=\;}
\def\+{\;+\;}
\def\Tr{{\rm Tr}}
\def\eps{\varepsilon}
\begin{document}
\begin{flushright}
August 2000
\end{flushright}
\vspace{1.0cm}
\begin{center}
\doublespace
\begin{large}
{\bf Meson loop effects in the NJL model at zero and non-zero temperature}\\
\end{large}
\vskip 1.0in
M. Oertel, M. Buballa and J. Wambach\\
{\small{\it Institut f\"ur Kernphysik, TU Darmstadt,\\ 
Schlossgartenstr. 9, 64289 Darmstadt, Germany}}\\
\end{center}
\vspace{1cm}

\begin{abstract}
We compare two different possibilities to include meson-loop corrections in
the Nambu--Jona-Lasinio model: a strict
$1/N_c$-expansion in next-to-leading order and a non-perturbative scheme
corresponding to a one-meson-loop 
approximation to the effective action.
Both schemes are consistent with chiral symmetry, in particular
with the Goldstone theorem and the Gell-Mann--Oakes--Renner relation. 
The numerical part at zero temperature focuses on the pion and the $\rho$-meson
sector. For the latter the meson-loop-corrections are crucial in order
to include the dominant $\rho \rightarrow \pi\pi$-decay channel,
while the standard Hartree + RPA approximation only contains unphysical 
$q\bar q$-decay channels. 
We find that $m_{\pi}, f_{\pi},\qq$ and quantities related to the
$\rho$-meson self-energy can be described
reasonably with one parameter set in the {\nce}, whereas we did not succeed to
obtain such a fit in the non-perturbative scheme.  
We also investigate the temperature dependence
of the quark condensate. Here we find consistency with chiral
perturbation theory to lowest order. Similarities
and differences of both schemes are discussed.

\end{abstract}

\newpage
\section{Introduction}
\singlespace
During the last few years one of the principal goals in nuclear
physics has been to explore the phase structure of QCD.
Along with this comes the investigation of hadron properties in
the vacuum as well as in hot or dense matter.
In principle, all properties of strongly interacting particles
should be derived from QCD. 
However, at least in the low-energy regime, where perturbation theory 
is not applicable, this is presently limited to a rather small number of 
observables which can be studied on the lattice, while more complex 
processes can either be addressed by chiral perturbation theory or within 
effective model calculations which try to incorporate the relevant
degrees of freedom.

So far the best descriptions of hadronic spectra, decays and
scattering processes are obtained within phenomenological hadronic models.
For instance the pion electromagnetic form factor in the time-like
region can be reproduced rather well within a simple vector dominance
model with a dressed $\rho$-meson which is constructed by coupling a
bare $\rho$-meson to a two-pion intermediate state \cite{brown,herrmann}. 
Models of this type have been successfully extended to investigate medium 
modifications of vector mesons and to calculate dilepton production rates
in hot and dense hadronic matter \cite{rapp}.     

In this situation one might ask how the phenomenologically successful
hadronic models emerge from the underlying quark structure and the
symmetry properties of QCD. Since this question cannot be answered 
at present from first principles it has to be addressed within quark 
models. For light hadrons chiral symmetry and its spontaneous breaking in 
the physical vacuum through instantons plays the decisive role in 
describing the two-point correlators \cite{schaefer} with 
confinement being much less important. This feature is 
captured by the Nambu--Jona-Lasinio (NJL) model in which the four-fermion
interactions can be viewed as being induced by instantons. 
Furthermore the model allows a study of the chiral phase transition as well as
the examination of the influence of (partial) chiral symmetry restoration on
the properties of light hadrons. 

The study of hadrons 
within the NJL model has of course a long history. In fact, mesons of various quantum numbers have already been discussed
in the original papers by Nambu and Jona-Lasinio \cite{njl} and by
many authors thereafter (for reviews see \cite{vogl,klevansky,hatsuda}). 

In most of these works quark masses are calculated in mean-field 
approximation  (Hartree or Hartree-Fock) while mesons are constructed as 
correlated quark-antiquark states (RPA). This corresponds to a leading-order 
approximation in $1/N_c$, the inverse number of colors. 
With the appropriate choice of parameters chiral symmetry, which is
an (approximate) symmetry of the model Lagrangian,
is spontaneously broken in the vacuum and pions emerge as (nearly) massless 
Goldstone bosons. While this is clearly one of the successes of the model, 
the description of other mesons is more problematic. 
One reason is the fact that the NJL model does not confine quarks.
As a consequence a meson can decay into free constituent quarks if its 
mass is larger
than twice the constituent quark mass $m$. Hence, for a typical value of
$m \sim$~300~MeV, the $\rho$-meson with a mass of 770~MeV, for
instance, would be unstable against decay into quarks.   
On the other hand the physical decay channel of the $\rho$-meson into two
pions is not included in the standard approximation. 
Hence, even if a large constituent quark mass is chosen in order to suppress 
the unphysical decays into quarks, one obtains a poor description of the 
$\rho$-meson propagator and related observables, like the pion electromagnetic 
form factor.

Similar problems arise if one wants to study the phase structure of
strongly interacting matter within a mean-field calculation for the NJL model,
although this has been done by many authors 
(see e.g.~\cite{klevansky,hatsuda,sklimt,mlutz}). 
In these calculations the thermodynamics is entirely driven by unphysical
unconfined quarks even at low temperatures and densities, whereas the 
physical degrees of freedom, in particular the pion, are missing. 

This and other reasons have motivated several authors to go beyond the 
standard approximation scheme and to include mesonic fluctuations.
In Ref.~\cite{krewald} a quark-antiquark $\rho$-meson is coupled via a quark
triangle to a two-pion state. 
Also higher-order corrections to the quark self-energy \cite{quack}
and to the quark condensate \cite{blaschke} have been investigated.
However, as the most important feature of the NJL model is chiral symmetry,
one should use an approximation scheme which conserves the symmetry 
properties, to ensure the existence of massless Goldstone bosons.

A non-perturbative symmetry conserving approximation
scheme has been discussed in Refs.~\cite{dmitrasinovic} and \cite{nikolov}. 
In Ref.~\cite{dmitrasinovic} a correction term to the quark self-energy
is included in the gap equation.
The authors find a consistent scheme to describe mesons and show the
validity of the Goldstone theorem and the Goldberger-Treiman relation
in that scheme. 
The authors of Ref.~\cite{nikolov} use a one-meson-loop approximation to the
effective action in a bosonized NJL model.
The structure of the meson propagators turns out to be the
same as in the approach of Ref.~\cite{dmitrasinovic}.
Based on this scheme various authors have investigated the effect of
meson-loop corrections on the pion electromagnetic form factor
\cite{lemmer} and on $\pi$-$\pi$ scattering in the vector \cite{he}
and the scalar channel \cite{huang}. 
However, since the numerical evaluation of the multi-loop diagrams is rather
involved, in these references the exact expressions are approximated by
low-momentum expansions. 

Another possibility to construct a symmetry conserving approximation scheme
beyond Hartree approximation and RPA is a strict $1/N_c$-expansion
up to next-to-leading order. 
Whereas in the approximation scheme mentioned above the gap equation is 
modified in a selfconsistent way, the corrections in the {\nce} are
perturbative. 
The consistency of the {\nce} with chiral symmetry has 
already been shown in Ref.~\cite{dmitrasinovic}. It has been studied in
more detail in
Refs.~\cite{oertel,OBW}. Recently such an expansion has been discussed also
in the framework of a non-local generalization of the NJL model~\cite{plant}.

In the present paper we compare the results obtained in the non-perturbative
scheme with those obtained in the {\nce}. In vacuum we focus our
discussion on the pion
and the $\rho$-meson, calculated with the full momentum dependence of all
expressions. Within the {\nce} the influence of mesonic
fluctuations on the pion propagator has been examined closely in
Ref.~\cite{oertel}. This was mainly motivated
by recent works by Kleinert and Van den Bossche \cite{kleinert}, who claim 
that chiral symmetry is {\it not} spontaneously broken in the NJL model as 
a result of strong mesonic fluctuations. In Ref.~\cite{oertel} we argue 
that because of the non-renormalizability of the NJL model new divergences 
and hence new cutoff parameters emerge if one includes meson loops.
Following Refs.~\cite{dmitrasinovic} and \cite{nikolov} we regularize the
meson loops by an independent cutoff parameter $\Lambda_M$. The results 
are, of course, strongly dependent on this parameter. Whereas for moderate 
values of $\Lambda_M$ the pion properties change only quantitatively, 
strong instabilities are encountered for larger values of $\Lambda_M$. In
Ref.~\cite{oertel} we suggested that this might 
be a hint for an instability of the spontaneously broken vacuum state. 
It turns out that the same type of instabilities also emerge
in the non-perturbative scheme. This allows for an analysis of the vacuum
structure and therefore for a more decisive answer to the question 
whether chiral symmetry gets indeed restored due to strong mesonic 
fluctuations within this approximation.

In any case, in the {\nce} the region of parameter values
where instabilities emerge in the pion propagator is far away from a
realistic parameter set~\cite{OBW}. 
We used $m_{\pi}, f_{\pi},\qq$ and the $\rho$-meson spectral function to fix
the parameters. The last one 
is particularly suited, as it cannot be described realistically without taking
into account pion loops, to fix the parameters. An important result of
the analysis in Ref.~\cite{OBW} was that such a fit can be
achieved with a constituent quark mass which is large enough such that the 
unphysical $q\bar q$-threshold opens above the $\rho$-meson peak. Since
the constituent quark mass is not an independent input parameter this
was not clear a priori. In this paper we will try the same for the
selfconsistent scheme. It turns out that it is not possible to find a
parameter set, where the constituent quark mass comes out large
enough to describe the properties of the $\rho$-meson
reasonably. In fact we encounter instabilities in the $\rho$-meson propagator
which are similar to those we found in the pion propagator for large
$\Lambda_M$. 

The inclusion of meson loop effects should also improve the thermodynamics of
the model considerably. A first insight on the influence of mesonic
fluctuations upon the thermodynamics can be obtained via the temperature
dependence of the quark condensate. It has been shown in
Ref.~\cite{florkowski} that in the selfconsistent scheme the low-temperature
behavior is dominated by pionic degrees of freedom which is a considerable
improvement on calculations in Hartree approximation where quarks
are the only degrees of freedom. Within this scheme the lowest-order
chiral perturbation theory result can be reproduced. This is also the case for
the {\nce} which will be demonstrated in the last part of this paper. 
The non-perturbative scheme also allows for an examination of the chiral phase
transition~\cite{florkowski}, whereas this is not possible within the {\nce}.

The paper is organized as follows.
In Sec.~\ref{model} we begin with a brief summary of the standard
approximation scheme used in the NJL model to describe quarks and mesons and afterwards
present the scheme for describing quantities
in next-to-leading order in $1/N_c$. In Sec.~\ref{nonpert} we discuss
the non-perturbative approximation scheme.
The consistency of these schemes
with the Goldstone theorem and with the Gell-Mann Oakes Renner relation
will be shown in Sec.~\ref{secpion}. 
The numerical results at zero temperature will be presented in
Sec.~\ref{numerics}. 
The temperature dependence of the quark condensate at non-zero temperature
within the two above mentioned approximation schemes will be studied in
\Sec{qqatt}. 
Finally, conclusions are drawn in Sec.~\ref{conclusions}.
\newpage


\section{The NJL model in leading order and next-to-leading order in $1/N_c$}
\label{model}
\setcounter{equation}{0}

\subsection{The standard approximation scheme: Hartree + RPA}
\label{model1}

We consider the following generalized NJL-model Lagrangian:
\beq
   {\cal L} \;=\; \pb ( i \partial{\hskip-2.0mm}/ - m_0) \psi
            \;+\; g_s\,[(\pb\psi)^2 + (\pb i\gamma_5{\vec\tau}\psi)^2]  
            \;-\; g_v\,[(\pb\gamma^\mu{\vec\tau}\psi)^2 + 
                        (\pb\gamma^\mu\gamma_5{\vec\tau}\psi)^2]   
   \,,
\eeq
where $\psi$ is a quark field with $N_f$~=~2 flavors and $N_c$~=~3 colors.
$g_s$ and $g_v$ are coupling constants with dimension $length^2$.
In contrast to QCD, color is not related to a gauge symmetry in this model, 
but only relates to a counting of degrees of freedom. However, if one 
defines the coupling constants to be of the order $1/N_c$, the large-$N_c$ 
behavior of the model agrees with that of QCD \cite{dmitrasinovic,nikolov}. 
Although we are not interested in the behavior of the model for arbitrary 
numbers of colors in the present article, the $1/N_c$-expansion is introduced 
for the purpose of book-keeping. This will allow us to take into account 
mesonic fluctuations in a symmetry conserving way.    
In order to establish the expansion scheme, the number of colors will be
formally treated as variable. All numerical calculations 
will be performed, however, with the physical value, $N_c$~=~3.

In the limit of vanishing current quark masses $m_0$ (``chiral limit'')
the above Lagrangian is invariant under global $SU(2)_L \times SU(2)_R$
transformations. For a sufficiently large scalar attraction this 
symmetry is spontaneously broken. This has mostly been studied within 
the (Bogoliubov-) Hartree approximation.\footnote{
Because of the local 4-fermion interaction in the Lagrangian,
exchange diagrams can always be cast in the form of direct diagrams via
a Fierz transformation. This is well known from zero-range interactions
in nuclear physics.  In particular the  Hartree-Fock approximation is 
equivalent to the Hartree approximation with appropriately redefined
coupling constants. 
In this sense we call the Hartree approximation the ``standard approximation''
to the NJL model, although in several references a Hartree-Fock approximation
has been performed.
}

\begin{figure}[b!]
\hspace{3cm}
\parbox{10cm}{
     \epsfig{file=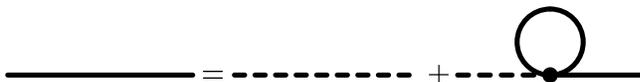}}
\caption{\it The Dyson equation for the quark propagator in the 
         Hartree approximation (solid lines). The dashed lines denote
         the bare quark propagator.}  
\label{fig1} 
\end{figure}

The Dyson equation for the quark propagator in Hartree approximation
is diagrammatically shown in \fig{fig1}.
The selfconsistent solution of this equation leads to a momentum 
independent quark self-energy $\Sigma_H$ and therefore only gives a correction 
to the quark mass:
\beq
m \= m_0 \+ \Sigma_H(m)
\= m_0 \+ \sum_M \; 2i g_M\ \intp \;\Tr\,[\,\Gamma_M\,S(p)\,]~.
\label{gap}
\eeq
Usually, $m$ is called the ``constituent quark mass''. 
Here $S(p) = (\psl - m)^{-1}$ is the (Hartree) quark propagator
and ``$\Tr$'' denotes a trace in color, flavor and Dirac space.
The sum runs over all interaction channels 
$M = \sigma, \pi, \rho, a_1$ with
$\Gamma_\sigma = 1\!\!1$, $\Gamma_\pi^a = i\gamma_5\tau^a$,
$\Gamma_\rho^{\mu\,a} = \gamma^\mu\tau^a$ and 
$\Gamma_{a_1}^{\mu\,a} = \gamma^\mu\gamma_5\tau^a$.
The corresponding coupling constants are $g_M = g_s$ for $M = \sigma$ or 
$M = \pi$ and $g_M = g_v$ for $M = \rho$ or $M = a_1$. Of course, only 
the scalar channel ($M = \sigma$) contributes in vacuum. One gets
\beq
m \= m_0 \+ 2i g_s\ 4 N_c N_f \intp {m\over{p^2-m^2+i\epsilon}}~.
\label{gapexp}
\eeq
In a $1/N_c$ expansion of the quark self-energy the Hartree approximation 
corresponds to the leading order.
Since $g_s$ is of the order $1/N_c$ the constituent quark mass $m$, and hence 
the quark propagator are of the order unity.

For sufficiently large couplings $g_s$ \eq{gapexp} allows for a finite 
constituent quark mass $m$ even in the chiral limit. 
In the mean-field approximation this solution minimizes the ground state 
energy. Because of the related gap in the quark spectrum, one usually refers 
to this equation as the gap equation, in analogy to BCS theory.

A closely related quantity is the quark condensate, which is generally
given by
\beq
    \qq \= -i \intp \; \Tr\,S(p) \;.
\label{qbq} 
\eeq
In Hartree approximation one immediately gets from the gap equation
\beq
    \qq^{(0)} \= - \frac{m-m_0}{2g_s} \;,
\label{qbq0}
\eeq     
where we have used the superscript $(0)$ to indicate that this corresponds
to a Hartree approximation.

\begin{figure}[t!]
\hspace{4cm}
\parbox{10cm}{
     \epsfig{file=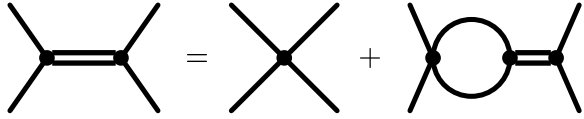}}
\caption{\it The Bethe-Salpeter equation for the meson propagator 
             in the RPA (double line). The solid lines indicate quark 
             propagators.}
\label{fig2} 
\end{figure}

Mesons are described via a Bethe-Salpeter equation. Here the leading order
in $1/N_c$ is given by a random phase approximation (RPA) without
Pauli-exchange diagrams. This is diagrammatically shown in \fig{fig2}. 
The elementary building blocks of this scheme are the quark-antiquark 
polarization functions
\beq
   \Pi_M(q)\= -i\intp \;\Tr\,[\,\Gamma_M \, iS(p+{q\over2})
                  \,\Gamma_M \, iS(p-{q\over2})\,] \;,
\label{pol0}
\eeq
with $\Gamma_M$, $M = \sigma, \pi, \rho, a_1$ as defined above.
Again, the trace has to be taken in color, flavor and Dirac space.
Iterating the scalar (pseudoscalar) part of the four-fermion 
interaction one obtains for the sigma meson (pion):
\beq
D_\sigma(q) \= \frac{-2 g_s}{1-2g_s\Pi_\sigma(q)} \;,\qquad
D_\pi^{ab}(q) \;\equiv\;D_\pi(q)\,\delta_{ab} 
\= \frac{-2 g_s}{1-2g_s\Pi_\pi(q)}\,\delta_{ab}\;.
\label{dsigmapi}
\eeq
Here $a$ and $b$ are isospin indices and we have used the notation
$\Pi_\pi^{ab}(q) \equiv \Pi_\pi(q)\,\delta_{ab}$.

In the vector channel this can be done in a similar way. Using the
transverse structure of the polarization loop in the vector channel,
\beq
    \Pi_\rho^{\mu\nu, ab}(q) \= \Pi_\rho(q)\,T^{\mu\nu}\, \delta_{ab}
    \;;\qquad T^{\mu\nu} =  (-g^{\mu\nu} + \frac{q^\mu q^\nu}{q^2})
    \;,
\label{pirho}
\eeq
one obtains for the $\rho$-meson
\beq
D_\rho^{\mu\nu,ab}(q) \;\equiv\; D_\rho(q)\;T^{\mu\nu}\;\delta_{ab}
\= \frac{-2 g_v}{1-2g_v\Pi_\rho(q)} \,T^{\mu\nu}\;\delta_{ab} 
 \;.
\label{drho}
\eeq
Analogously, the $a_1$ can be constructed from the transverse part of 
the axial polarization function $\Pi_{a_1}$.
As discussed e.g. in Ref.~\cite{klimt} $\Pi_{a_1}^{\mu\nu}$ also contains 
a longitudinal part which contributes to the pion. Although there is no 
conceptional problem in including this mixing we will neglect it in the 
present paper in order to keep the structure of the model as simple as 
possible.

It follows from Eqs.~(\ref{pol0})~-~(\ref{drho}) that the functions
$D_M(q)$ are of order $1/N_c$. Their explicit forms are given
in App.~\ref{correlators}. For simplicity we will call them
``propagators'', although strictly speaking, they should be interpreted as 
the product of a renormalized meson propagator with a squared 
quark-meson coupling constant.
The latter is given by the inverse residue of the function $D_M(q)$,
while the pole position determines the meson mass:
\beq
   D_M^{-1}(q)|_{q^2 = m_M^{2 (0)}} \= 0 \;,\qquad
    g_{Mqq}^{-2 (0)} \= \frac{d\Pi_M(q)}{dq^2}|_{q^2 = m_M^{2 (0)}} \;.
\label{mesonmass0}
\eeq
Again the superscript $(0)$ indicates that $m_M^{2 (0)}$
and $g_{Mqq}^{-2 (0)}$ are quantities in RPA. 
One easily verifies that they are of order unity and $1/\sqrt{N_c}$,
respectively.


\subsection{Next-to-leading order corrections}
\label{model2}

With the help of the gap equation, Eq.~(\ref{gapexp}), one can show that 
the ``standard scheme'', i.e. Hartree approximation + RPA, is 
consistent with chiral symmetry. For instance, in the chiral limit
pions are massless, as required by the Goldstone theorem. 
Of course one would like to preserve this feature when one goes beyond 
the standard scheme. One way to accomplish this is to perform a strict
$1/N_c$ expansion, systematically including higher-order corrections. 
In this subsection we want to construct the quark self-energy and the
mesonic polarization functions in next-to-leading order in $1/N_c$.

\begin{figure}[t!]
\hspace{5cm}
\parbox{6cm}{
     \epsfig{file=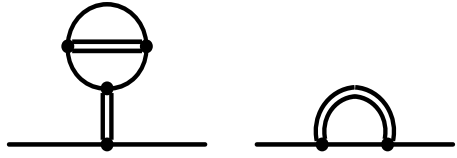,
     height=2.cm, width=6.0cm}}
\caption{\it The $1/N_c$-corrections $\delta\Sigma^{(a)}$ (left) and 
$\delta\Sigma^{(b)}$ (right) to the quark self-energy.}
\label{fig3} 
\end{figure}

The correction terms to the quark self-energy, 
\beq
    \delta \Sigma(p) \= \delta\Sigma^{(a)} \+ \delta\Sigma^{(b)}(p) \;,
\eeq
are shown in \fig{fig3}. 
In these diagrams the single lines and the double lines
correspond to quark propagators in the Hartree approximation (order unity)
and to meson propagators in the RPA (order $1/N_c$), respectively. 
Recalling that one obtains a factor $N_c$ for a closed quark loop one
finds that both diagrams are of order the $1/N_c$. One can also easily
convince oneself that there are no other self-energy diagrams of that order.  

According to \eq{qbq}, the $1/N_c$-correction to the quark condensate
is given by
\beq
    \delta\qq \= -i \intp \; \Tr\,\delta S(p) \;,
\label{deltaqbq} 
\eeq
with
\beq 
    \delta S(p) \= S(p)\,\delta\Sigma(p)\,S(p) 
\label{deltaS}
\eeq
being the $1/N_c$-correction to the Hartree quark propagator $S(p)$. 
Since we are interested in a strict $1/N_c$ expansion, the self-energy
correction must not be iterated.

\begin{figure}[t!]
\hspace{1cm}
\parbox{14cm}{
     \epsfig{file=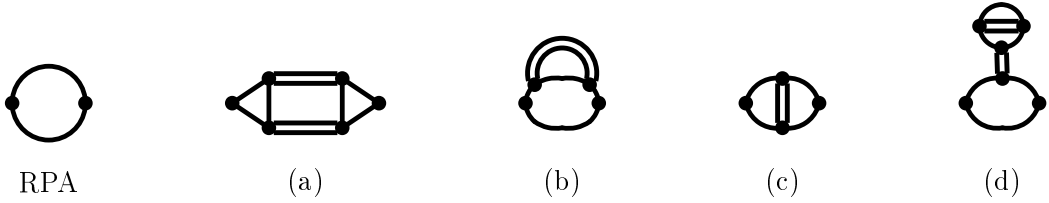,
     height=2.6cm, width=14.0cm}}
\caption{\it Contributions to the mesonic polarization function
             in leading (RPA) and next-to-leading order in $1/N_c$.}
\label{fig4} 
\end{figure}

The $1/N_c$-corrected mesonic polarization diagrams read
\beq
    {\tilde \Pi}_M(q) \= \Pi_M(q) \+ \sum_{k=a,b,c,d}\; \delta \Pi_M^{(k)}(q)
\;.
\label{pol1}
\eeq
The four correction terms $\delta \Pi_M^{(a)}$ to $\delta \Pi_M^{(d)}$
together with the leading-order term $\Pi_M$ are shown in \fig{fig4}. 
Again, the lines in this figure correspond to Hartree quarks and RPA mesons.
Since the correction terms consist of either one RPA propagator and one
quark loop or two RPA propagators and two quark loops they are of the
order unity, whereas the leading-order term is of the order $N_c$. 

In analogy to Eqs.~(\ref{dsigmapi}), (\ref{drho}) and (\ref{mesonmass0})
the corrected meson propagators are given by
\beq
{\tilde D}_M(q) \= \frac{-2 g_M}{1-2g_M {\tilde\Pi}_{M}(q)} \;,
\label{dm1}
\eeq
while the corrected meson masses are defined by the pole positions 
of the propagators:
\beq
   \tilde{D}_M^{-1}(q)|_{q^2 = m_M^2} \= 0 \;.
\label{mesonmass1}
\eeq
As we will see in \Sec{piondstl} this scheme is consistent with the
Goldstone theorem, i.e. in the chiral limit it leads to massless pions.
Note, however, that because of its implicit definition $m_M$ contains terms 
of arbitrary orders in $1/N_c$, although we start from a strict expansion of
the inverse meson propagator up to next-to-leading order.
This will be important in the context of the Gell-Mann--Oakes--Renner relation.

\begin{figure}[b!]
\hspace{3cm}
\parbox{10cm}{
     \epsfig{file=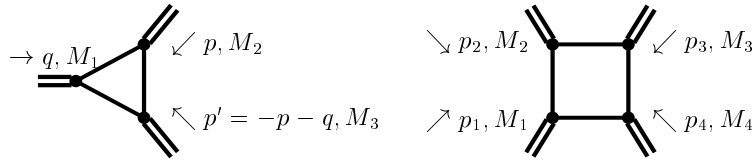,
     height=2.1cm, width=10.cm}}
\caption{\it (Left) The quark triangle vertex $ -i\Gamma_{M_1,M_2,M_3}(q,p)$.
             (Right) The quark box vertex
              $-i\Gamma_{M_1,M_2,M_3,M_4}(p_1,p_2,p_3)$.}
\label{fig5} 
\end{figure}

For a more explicit evaluation of the correction terms it is advantageous
to introduce the quark triangle and box diagrams which are shown in
\fig{fig5}.
The triangle diagrams entering into $\delta\Sigma^{(a)}$, 
$\delta \Pi_M^{(a)}$ and $\delta \Pi_M^{(d)}$ can be interpreted as 
effective three-meson vertices.
For external mesons $M_1$, $M_2$ and $M_3$ they are given by
\bea
 -i \Gamma_{M_1,M_2,M_3}(q,p) &=& - \intk
  \Big\{ \Tr\,[\Gamma_{M_1}i S(k)\Gamma_{M_2} i S(k-p)\Gamma_{M_3}
 i  S(k+q)] \nonumber \\
& & \hspace{1.85cm}+ \Tr\,[\Gamma_{M_1}i S(k-q)\Gamma_{M_3}i 
    S(k+p)\Gamma_{M_2} i S(k)]\Big\}~, 
\label{trianglevertex}
\eea
with the operators $\Gamma_M$ as defined below Eq.~(\ref{gap}).
We have summed over both possible orientations of the quark loop.
For later convenience we also define the constant
\beq
    \Delta \= \frac{1}{2}\intp\sum_M (-iD_M(p))
 (-i\Gamma_{M,M,\sigma}(p,-p)) \;, 
\label{Delta}
\eeq
which corresponds to a quark triangle coupled to an external scalar vertex
and a closed meson loop. 

The quark box diagrams are effective four-meson vertices and are needed 
for the evaluation of $\delta \Pi_M^{(b)}$ and $\delta \Pi_M^{(c)}$.
If one again sums over both orientations of the quark loop they are given
by  
\bea
&&  \hspace{-2.0cm}
-i\Gamma_{M_1,M_2,M_3,M_4}(p_1,p_2,p_3) \phantom{\intk}\nonumber\\
&=&\intk \Big(
\Tr\,[\Gamma_{M_1}iS(k)\Gamma_{M_2}iS(k-p_2)\Gamma_{M_3}iS(k-p_2-p_3)
    \Gamma_{M_4}iS(k+p_1)]  
\nonumber\\ && \hspace{1.4cm}
 +\Tr\,[\Gamma_{M_1}iS(k-p_1)\Gamma_{M_4}iS(k+p_2+p_3)\Gamma_{M_3}iS(k+p_2)
      \Gamma_{M_2}iS(k)]\Big)~.
\label{boxvertex}
\eea

With these definitions the various diagrams can be written in a relatively 
compact form. 
For the momentum independent correction term to the quark self-energy we get
\beq 
     \delta\Sigma^{(a)} \= - \frac{1}{2} D_\sigma(0) \; 
     \sum_{M}\;\intp \;D_M(p)\,\Gamma_{M,M,\sigma}(p,-p) 
     \= D_\sigma(0)\,\Delta \;.
     \label{deltasigmaa}
\eeq
In principle there should be also a sum over the quantum numbers of the 
meson which connects the quark loop with the external quark legs, but
all contributions from other mesons than the $\sigma$-meson vanish.
The factor of $1/2$ is a symmetry factor which is needed because otherwise 
the sum over the two orientations of the quark propagators,
which is contained in the definition of the quark triangle vertex 
(Eq.~(\ref{trianglevertex})) would lead to double counting.

The evaluation of the momentum dependent correction term 
$\delta\Sigma^{(b)}$ is straight forward:
\beq 
     \delta\Sigma^{(b)}(k) \=  i \sum_{M}\; \intp \; D_{M}(p)\;
     \Gamma_M\,S(k-p)\,\Gamma_M \;.
     \label{deltasigmab}
\eeq
Inserting these expressions for $\delta\Sigma^{(a)}$ and $\delta\Sigma^{(b)}$ 
into \eq{deltaS}, the $1/N_c$-correction term to the quark condensate, 
\eq{deltaqbq}, can be brought into the form
\beq
    \delta\qq \= - \frac{D_\sigma(0)\,\Delta}{2 g_s}\;.
\label{deltaqbqexp}
\eeq

For the mesonic polarization diagrams we get:
\bea
\delta\Pi^{(a)}_{M}(q) &\=& \phantom{-} \frac{i}{2}\intp \sum_{M_1 M_2}
\Gamma_{M,M_1,M_2}(q,p)\, D_{M_1}(p)\,\Gamma_{M,M_1,M_2}(-q,-p)\,
D_{M_2}(-p-q)
\;, \nonumber\\
\delta\Pi^{(b)}_{M}(q)&\=&- i\,\intp \hspace{3.0mm} \sum_{M_1} \;
\Gamma_{M,M_1,M_1,M}(q,p,-p)\,D_{M_1}(p)
\;, \nonumber\\
\delta\Pi^{(c)}_M(q)&\=& -\frac{i}{2}\intp \hspace{2.5mm} \sum_{M_1} \; 
\Gamma_{M,M_1,M,M_1}(q,p,-q)\,D_{M_1}(p)
\;, \nonumber\\
\delta\Pi^{(d)}_M(q) &\=& \frac{i}{2}\;\Gamma_{M,M,\sigma}(q,-q)\, D_\sigma(0)
\; \intp \hspace{3.0mm} \sum_{M_1}\; 
\Gamma_{M_1,M_1,\sigma}(p,-p) \, D_{M_1}(p)
\;, \nonumber\\
&\=& -i\,\Gamma_{M,M,\sigma}(q,-q)\,D_\sigma(0)\,\Delta
\;.  
\label{deltapi}
\eea
The symmetry factor of $1/2$ for $\delta\Pi^{(c)}_M$ and $\delta\Pi^{(d)}_M$ 
has the same origin as in \eq{deltasigmaa}.
Similarly in $\delta\Pi^{(a)}_M$ we had to correct for the fact that the
exchange of $M_1$ and $M_2$ leads to identical diagrams.

For the further evaluation of Eqs.~(\ref{deltasigmaa}) to (\ref{deltapi}) we 
proceed in two steps. 
In the first step we calculate the intermediate RPA meson-propagators.
Simultaneously we can calculate the quark triangles and box diagrams.
One is then left with a meson loop which has to be evaluated in a second step. 

The various sums in Eqs.~(\ref{deltasigmaa}) to (\ref{deltapi}) are, 
in principle, over all quantum numbers of the intermediate mesons. 
However, for most applications we expect that the most important 
contributions come from the pion, which is the lightest particle in the game.
For instance, the change of the quark condensate at low temperatures
should be dominated by thermally excited pions.
Also, for a proper description of the $\rho$-meson width in vacuum we only
need the two-pion intermediate state in diagram $\delta\Pi^{(a)}_M$. 
Other contributions to this diagram, i.e $\pi a_1$, $\rho\sigma$, $\rho\rho$ 
and $a_1 a_1$ intermediate states, are much less important since the
corresponding decay channels open far above the $\rho$-meson mass and
- in the NJL model - also above the unphysical two-quark threshold.
Hence, from a purely phenomenological point of view, it should be sufficient 
for many applications
to restrict the sums in Eq.~(\ref{deltapi}) to intermediate pions.
However, in order to stay consistent with chiral symmetry, we have to 
include intermediate sigma mesons as well. On the other hand,  
vector- and axial-vector mesons can be neglected without violating chiral 
symmetry.
Since this leads to an appreciable simplification of the numerics
we have restricted the {\it intermediate} degrees of freedom to 
scalar and pseudoscalar mesons in the present paper.   
Of course, in order to describe a $\rho$-meson, we have to take vector
couplings at the external vertices of the diagrams shown in \fig{fig4}.


\section{Non-perturbative symmetry conserving schemes}
\label{nonpert}
\setcounter{equation}{0}

\subsection{Axial Ward identities}
\label{nonpert1}

The disadvantage of the $1/N_c$-expansion scheme is that it is perturbative.
Although we have constructed the $1/N_c$ corrections to the Hartree
quark self-energy (\fig{fig3}) we did not selfconsistently include
such diagrams in the gap equation. 
Since the iteration would produce terms of arbitrary orders in $1/N_c$,
one is not allowed to do so in a strict expansion scheme.
Therefore, all correction diagrams we have discussed in the previous section
consist of ``Hartree'' quark propagators. 
This perturbative treatment should work rather well as long as the 
$1/N_c$ corrections to the quark self-energy are small compared with the 
leading order, i.e. the constituent quark mass.
On the other hand it is clear that the scheme must fail to describe
the chiral phase transition, e.g. at finite temperatures. 
Here a non-perturbative treatment is mandatory. 

Therefore, in this section, we want to follow a different strategy,
exploiting the fact that the Goldstone theorem is basically a 
consequence of Ward identities: 
Consider an external axial current $j_{\mu 5}^a$ coupled to a quark. 
Then, in the chiral limit, the corresponding 
vertex function $\Gamma_{\mu 5}^a$ is related to the quark propagator 
$S(p)$ via the axial Ward-Takahashi identity
\beq
    q^\mu\,\Gamma_{\mu 5}^a(p,q) \;=\; S^{-1}(p+q)\,\gamma_5 \tau^a \;+\;
    \gamma_5 \tau^a \, S^{-1}(p),
\label{AWT}
\eeq 
where $p$ and $p+q$ are the 4-momenta of the incoming and outgoing quark, 
respectively. 
Obviously, for a non-vanishing constituent quark mass, the
r.h.s. of this equation remains finite even for $q \rightarrow 0$.
Consequently $\Gamma_{\mu 5}^a(p,q)$ must have a pole in this limit,
which can be identified with the Goldstone boson. Moreover the
explicit structure of the Goldstone boson can be constructed from
the structure of the axial vertex function. 

\begin{figure}[t!]
\hspace{1cm}
\parbox{16cm}{
     \epsfig{file=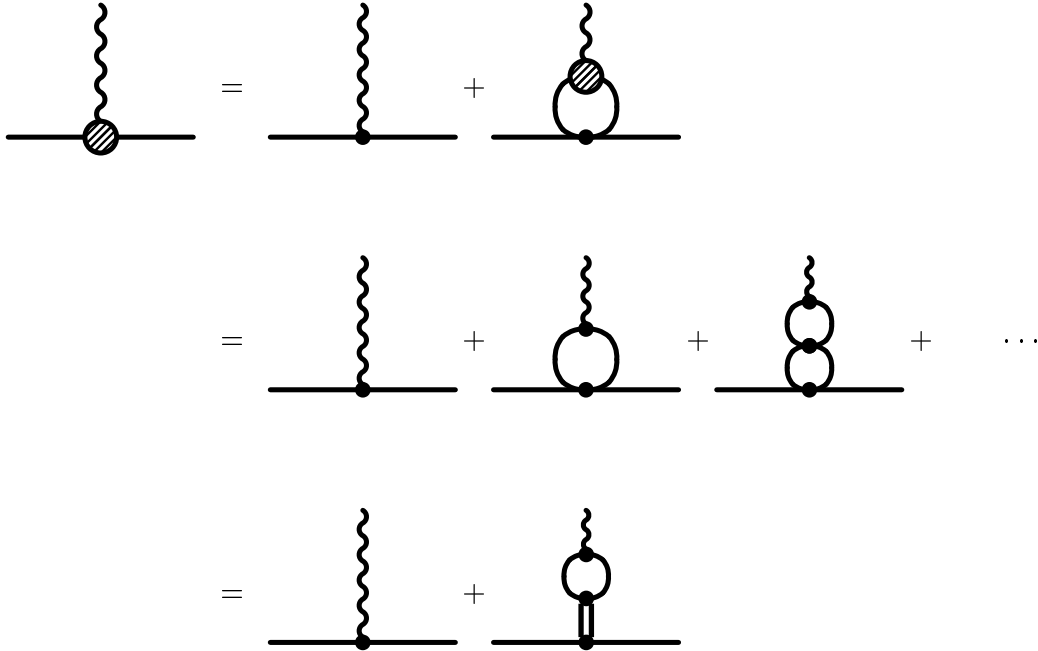}}
\caption{\it Vertex function for an external axial current coupled to
             a ``Hartree'' quark.}  
\label{fig6} 
\end{figure}

As a first example, let us start again from the Hartree gap equation 
(\eq{gap}, \fig{fig1}) and construct the axial vertex function by coupling the 
propagator to an external axial current. This is illustrated in 
\fig{fig6}. In the upper line, the first term on the r.h.s. describes 
the coupling to the bare quark,
corresponding to a bare vertex $\gamma_\mu\gamma_5\tau^a$. 
In the second term, however, the current is coupled to a dressed
quark, and therefore we have to use the same vertex function as on the 
l.h.s. of the equation: 
\beq
    \Gamma_{\mu 5}^{a}(p,q) \;=\;  \gamma_\mu\gamma_5\tau^a \;+\;
    \sum_M 2ig_M\, \Gamma_M \, \intk \,
    \Tr\,[\,\Gamma_M\,S(k+q)\,\Gamma_{\mu 5}^{a}(k,q) \,S(k)\,] \;.
\label{G50}
\eeq
Here $S(k)$ denotes the quark propagator in the Hartree approximation. 
As in \eq{gap}  the sum runs over all interaction channels, but of course 
only the pseudoscalar and the axial vector contributions do not vanish. 
Contracting \eq{G50} with $q_\mu$ one obtains a linear equation for
$q_\mu\,\Gamma_{\mu 5}^a$. One can easily verify that in the chiral limit
the solution of this equation is given by the axial Ward-Takahashi 
identity, \eq{AWT}. 
To this end we replace $q_\mu\,\Gamma_{\mu 5}^a$ on both sides of the
equation by the expressions given by \eq{AWT} and check whether the
results agree. On the r.h.s. the insertion of \eq{AWT}
basically amounts to removing one of the quark propagators from the loop.
In this way the loop receives the structure of the quark self-energy and
we can use the gap equation, \eq{gap} to simplify the expression.
For $m_0$~=~0 the result turns out to be equal to the l.h.s. of the
equation, which proves the validity of the axial Ward-Takahashi in this
scheme.
 
We have seen above, that this implies the existence of a massless Goldstone 
boson in the chiral limit. As illustrated in the second and third line of 
\fig{fig6}, the selfconsistent structure of \eq{G50} for the dressed vertex
$\Gamma_{\mu 5}^a$ leads to an iteration of the quark loop and an RPA pion 
emerges.
Hence we can identify the Goldstone boson with an RPA pion. 

Obviously the above procedure can be generalized to other cases: 
Starting from any given gap equation for the quark propagator we 
construct the vertex function to an external axial current by coupling 
the current in all possible ways to the r.h.s. of the equation. 
As long as the gap equation does not violate chiral symmetry
this automatically guarantees the validity of the axial Ward-Takahashi 
identity and therefore the existence of a massless pion in the chiral 
limit. The structure of this pion can then be obtained from the 
structure of the vertex correction.

\begin{figure}[b!]
\hspace{0cm}
\parbox{16cm}{
     \epsfig{file=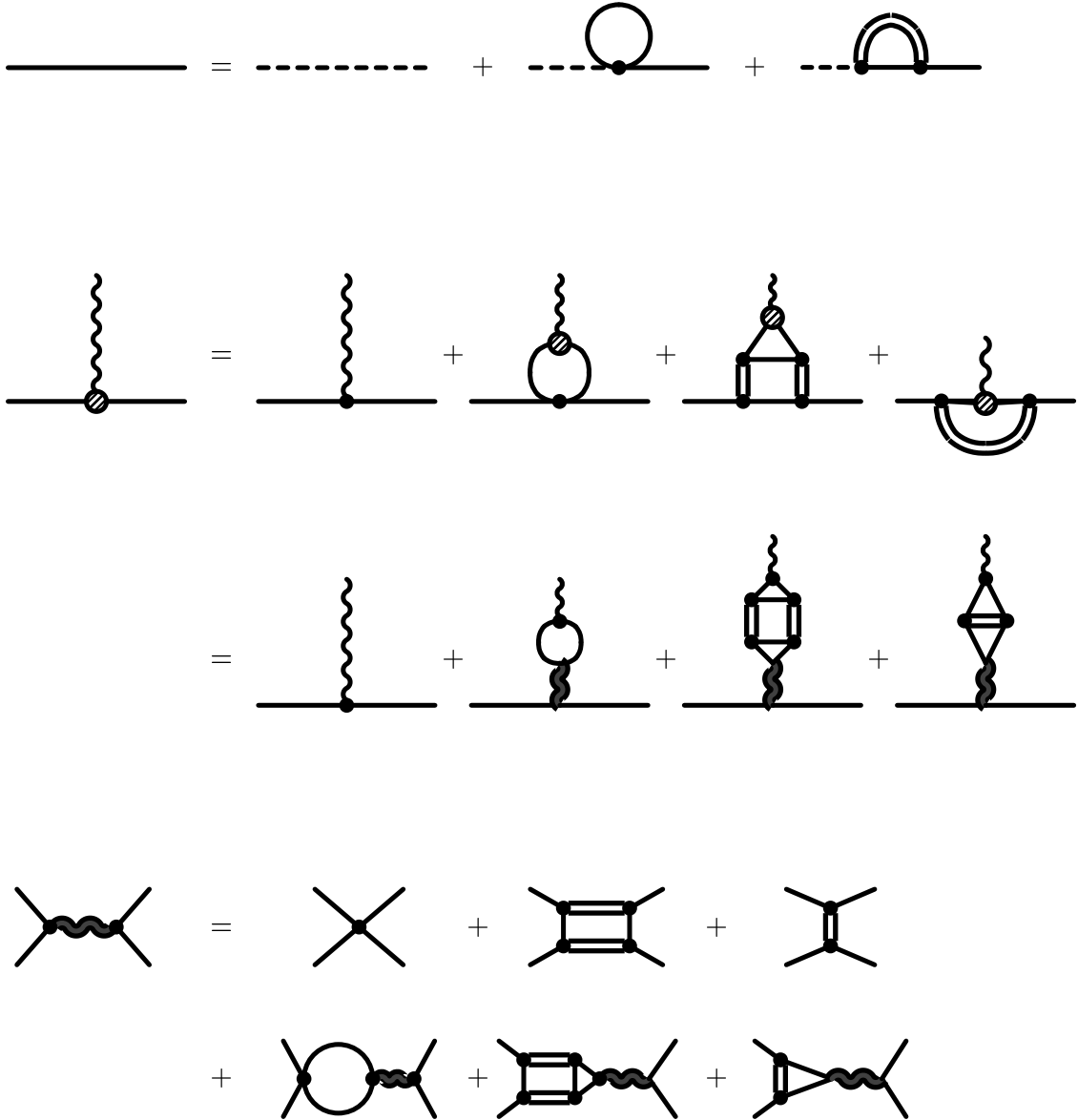}}
\caption{\it Selfconsistent scheme with a non-local self-energy term:
             Gap equation (upper part), equation for the vertex function 
             of an external current (middle) and the corresponding 
             equation for the quark-antiquark T-Matrix (lower part). 
             The double line denotes an RPA meson propagator 
             (see \fig{fig2}), which is selfconsistently constructed from 
             the dressed quark propagators of the present equation
             (solid line).} 
\label{fig7} 
\end{figure}

As an example we start from the extended gap equation depicted in the
upper part of \fig{fig7}. There, in addition to the Hartree term, the 
quark is dressed by RPA mesons. These are defined 
in the same way as before (\fig{fig2}), but now selfconsistently using 
the quark propagator which results from the extended gap equation. 
Therefore the RPA pions are no longer
massless in the chiral limit. However, following the strategy described
above we can construct the consistent pion propagator. To that end
we couple again an external axial current to both sides of the gap equation. 
The resulting equation for the vertex function is also shown in \fig{fig7}
(middle part). The additional term in the gap equation leads to two new 
diagrams which were not present in \fig{fig6}: In the first the current 
couples to 
a quark-antiquark loop of the RPA meson while in the second it couples to 
the quark inside of the meson loop. Again, one can easily check that the 
vertex function and the quark propagator fulfill the axial 
Ward-Takahashi identity \eq{AWT} in the chiral limit. 

In principle one can construct the corresponding massless Goldstone 
boson from the quark-antiquark T-matrix given in the lower part of \fig{fig7}. 
In practice, however, this is very difficult. In fact, already the solution
of the extended gap equation is difficult, since the additional
self-energy term is non-local, leading to a non-trivial 4-momentum dependence
of the quark propagator. Note that this propagator has to be selfconsistently
used for the calculation of the RPA-meson propagator. 
Therefore the authors of Ref.~\cite{dmitrasinovic} suggested to drop the 
non-local terms, but to keep a particular class of local diagrams which 
arises from the combined iteration of the quark loop and the meson loop.
This gap equation is shown in \fig{fig8}.
Because of the restriction to local self-energy insertions, we will call
this scheme the ``local selfconsistent scheme'' (LSS).  
It will be discussed in the next subsection.


\subsection{The local selfconsistent scheme}
\label{nonpert2}

\begin{figure}[b!]
\hspace{0cm}
\parbox{16cm}{
     \epsfig{file=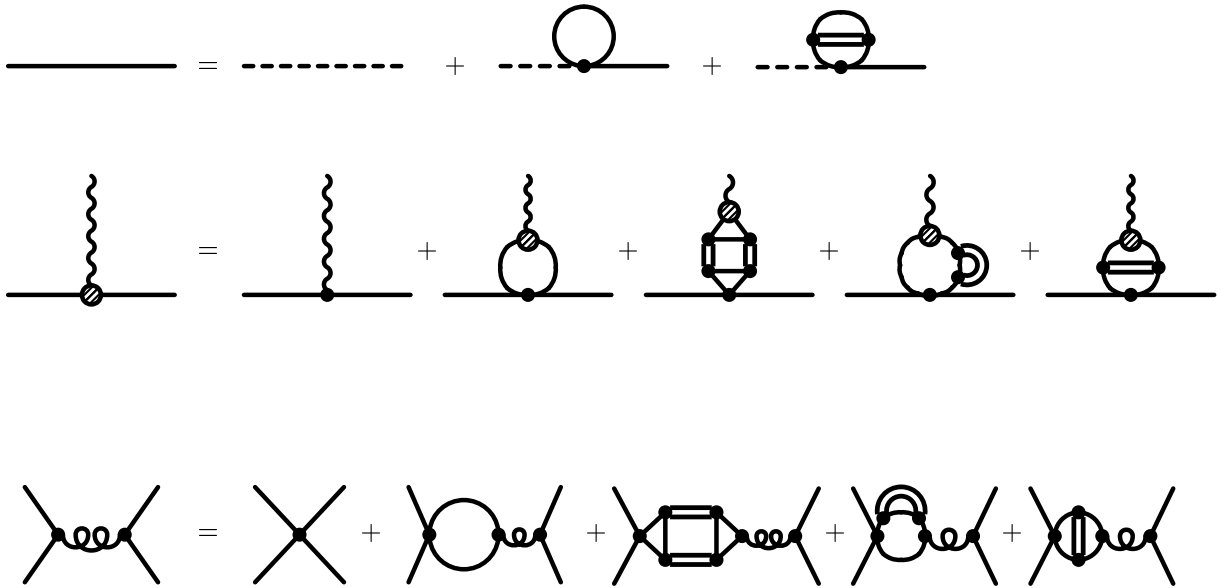,width=16cm,height=7.72cm}}
\caption{\it The ``local selfconsistent scheme'': Gap equation (upper part),
             equation for the vertex function of an external current 
             (middle) and the corresponding equation for the 
             consistent meson propagator (lower part). 
             The double line denotes an RPA meson propagator 
             (see \fig{fig2}), which is selfconsistently constructed from 
             the dressed quark propagators of the present equation
             (solid line).}   
\label{fig8} 
\end{figure}      

The gap equation for the constituent quark mass in the local selfconsistent 
scheme (upper part of \fig{fig8}) reads
\beq
    m \= m_0 \+ \tilde\Sigma(m) \= m_0 \+ \Sigma_H(m) 
             \+ \delta\tilde\Sigma(m)
\;.
\label{localgap}
\eeq
Here $\Sigma_H$ is the Hartree contribution to the self-energy as defined
in \eq{gap}. The correction term $\delta\tilde\Sigma$ 
corresponds to the third diagram on the r.h.s. of \fig{fig8}.
We have explicitly indicated that the self-energy diagrams have to be 
evaluated selfconsistently at the quark mass $m$, which comes out
of the equation. Because of the new diagram $\delta\tilde\Sigma$,
this mass is in general different from the Hartree mass.
However, since all diagrams in the LSS are 
constructed from the constituent quarks of \eq{localgap}, we prefer
not to introduce a new symbol for this mass.
This has the advantage that we can also keep the notations for the
quark propagator $S(p) = (\psl - m)^{-1}$, quark-antiquark loops,
triangles etc. which we introduced earlier. The general structure of
these diagrams is the same in all schemes we discuss in this article. 
Therefore we introduce the convention that in the $1/N_c$-expansion
scheme $m$ denotes the Hartree mass, while it denotes the solution of
\eq{localgap} in the LSS and all diagrams
should be evaluated at that mass, unless stated otherwise.

The self-energy term $\delta\tilde\Sigma$ consists of a quark loop dressed 
by an RPA-meson loop.
The quark loop is coupled to the external quark propagators via the
NJL point interaction. It can again be shown, that only the scalar
interaction contributes. 
Hence $\delta\tilde\Sigma$ is given by
\beq
    \delta\tilde\Sigma \= -2g_s\,\Delta   \;,
\label{deltatildesigma}
\eeq
where $\Delta$ is the constant defined in \eq{Delta}.

Because of this additional self-energy diagram in the gap equation,
the RPA is not the consistent scheme to describe mesons:
In the chiral limit, RPA pions are no longer massless. Hence, 
in order to find the consistent meson propagators, we proceed in
the way discussed in the previous subsection. 

The equation for the axial vertex function is shown in the middle part of 
\fig{fig8}. Compared with the corresponding equation which follows from 
the Hartree approximation (\fig{fig6}) there are three extra terms.
This leads to three additional polarization diagrams, which have to
be iterated in the Bethe-Salpeter equation for the consistent meson
propagator (lower part of \fig{fig8}). 

Obviously these diagrams are identical to $\delta\Pi^{(a)}_{M}$, 
$\delta\Pi^{(b)}_{M}$ and $\delta\Pi^{(c)}_{M}$, which we defined in
Sec.~\ref{model2} (\fig{fig4}, \eq{deltapi}), i.e. the new meson
propagators are given by 
\beq
{\tilde D}_M(q) \= \frac{-2 g_M}{1-2g_M {\tilde\Pi}_{M}(q)} \;,
\label{dmtilde}
\eeq
with
\beq
    {\tilde \Pi}_M(q) \= \Pi_M(q) \+ \sum_{k=a,b,c}\; \delta \Pi_M^{(k)}(q)
\;.
\label{poltilde}
\eeq

This structure agrees with the result of Ref.~\cite{dmitrasinovic}. 
In that reference the scheme was motivated by a $1/N_c$ expansion. 
However, one should stress again, that the selfconsistent solution
of the gap equation mixes all orders in $1/N_c$. Moreover,
the next-to-leading order self-energy correction term $\delta\Sigma^{(b)}$
(cf. \fig{fig3}) is not contained in the gap equation of \fig{fig8}.
Therefore the consistency of the scheme cannot be explained 
by $1/N_c$-arguments. In fact, our discussion shows that the structure 
of the consistent pion propagator can be derived from the gap equation 
without any reference to $1/N_c$ counting. 

On the other hand, if one performs a strict $1/N_c$ expansion of the 
mesonic polarization diagrams up to next-to-leading order one exactly 
recovers the diagrams shown in \fig{fig4} \cite{dmitrasinovic}.
This is quite obvious for the diagrams $\delta\Pi_M^{(a)}$ to
$\delta\Pi_M^{(c)}$, which are explicitly contained in \eq{poltilde}.
Diagram $\delta\Pi_M^{(d)}$, which seems to be missing, is implicitly
contained in the quark-antiquark loop via the next-to-leading order
terms in the quark propagator, which arise from the extended gap equation. 
 
In this sense, the LSS may be viewed as the
simplest non-perturbative extension of the standard scheme which is
consistent with the Goldstone theorem and which contains all mesonic
polarization diagrams up to next-to-leading order in $1/N_c$.
However, since the diagram $\delta\Sigma^{(b)}$ is not contained in
the gap equation, this is not true for the quark condensate: 
If we evaluate \eq{qbq} with the quark propagator of the present scheme,
we obtain
\beq
    \qq \= - \frac{\Sigma_H}{2g_s} \= - \frac{m - m_0}{2 g_s} \;-\; \Delta
    \;.
\label{qbqself1}
\eeq
Performing a strict $1/N_c$ expansion of this expression and only keeping
the next-to-leading order term one does not recover \eq{deltaqbqexp} but only 
the contribution of $\delta\Sigma^{(a)}$. 
This might be the reason why the authors of Ref.~\cite{dmitrasinovic}
determine the quark condensate as
\beq
    \qq \= - \frac{\tilde\Sigma}{2g_s} \= - \frac{m - m_0}{2 g_s}
    \;.
\label{qbqself2}
\eeq
In contrast to \eq{qbqself1} this expression reduces to the perturbative
result in a strict $1/N_c$ expansion. Moreover, as we will discuss in
\Sec{piondstl}, it is consistent with the Gell-Mann--Oakes--Renner relation.
On the other hand, \eq{qbqself2} does obviously not follow from \eq{qbq} 
with the quark propagator of the present scheme. 
A possible resolution to this problem was given in Ref.~\cite{nikolov},
where the LSS was derived using functional methods.
The meson propagators obtained in that way are identical to 
Eqs.~(\ref{dmtilde}) and (\ref{poltilde}), while the quark condensate is 
given by \eq{qbqself2}. 
This will be briefly discussed in the following subsection. 

Finally, we would like to comment on the name ``local selfconsistent
scheme'', which we have introduced in order to distinguish this scheme
from the perturbative $1/N_c$ expansion. We call this scheme
``selfconsistent'' because the quark propagator, which is determined
by the gap equation is selfconsistently used in the loops and the
RPA-meson propagator on the r.h.s. of that equation. However, as we
have seen, the scheme is not selfconsistent with respect to the mesons:
The improved meson propagators given by Eqs.~(\ref{dmtilde}) and 
(\ref{poltilde}) are different from the RPA mesons which are used in
the gap equation and hence as intermediate states in the mesonic
polarization functions $\delta\Sigma^{(a)}$ to $\delta\Sigma^{(c)}$.
On the other hand, if we had used the improved mesons already in the gap 
equation, our method of \Sec{nonpert1} would have led to further mesonic
polarization diagrams in order to be consistent
with chiral symmetry. Obviously, the construction of an expansion scheme,
which is selfconsistent for quarks and mesons, is an extremely difficult
task.


\subsection{One-meson-loop order in the effective action formalism}
\label{EAF}

Both, the non-local selfconsistent scheme,
which we briefly discussed in \Sec{nonpert1} (\fig{fig7}), and the local
selfconsistent scheme can be derived from
functional methods: The non-local selfconsistent scheme can be obtained as
a $\Phi$-derivable theory \cite{luttinger,baym} if one includes the
``ring sum'' in the generating functional. 
The present section is devoted to a brief discussion on how the local selfconsistent
scheme can be derived from a one-meson-loop approximation to the effective
action. The interested reader is referred to Refs.~\cite{nikolov,ripka}. 
Here we will basically follow Ref.~\cite{nikolov}. 

In this section we drop the vector and axial vector interaction and
start from a Lagrangian which contains only scalar and pseudoscalar
interaction terms:
\beq
   {\cal L} \;=\; \pb ( i \partial{\hskip-2.0mm}/ - m_0) \psi
            \;+\; g_s\,[(\pb\psi)^2 + (\pb i\gamma_5{\vec\tau}\psi)^2]  
   \,.
\eeq
The partition function of the system can be expressed in terms of the path
integral 
\beq
 Z= e^{-W} = \int {\mathcal D}(\pd){\mathcal D}(\psi)
 e^{-I(\pd,\psi)}~,
\label{z}
\eeq
with the Euclidean action 
\beq
I(\pd,\psi) = \int d^4 x_E \;\Big\{\pd \gamma_0 (\partial_{\tau}\gamma_0 
- i\vec{\gamma}\cdot\vec{\nabla} + m_0) \psi - g_s ( (\pd\gamma_0\psi)^2 
+ (\pd \gamma_0
i\gamma_5\vec{\tau}\psi)^2 )\Big\}~.
\eeq
The integration is here over a Euclidean space-time volume $d^4 x_E$, 
where $\partial_{\tau}$ corresponds to $i \partial_t$.
The standard procedure is now to bosonize the action by introducing auxiliary
fields $\Phi_a^{\prime}, a = \{0,1,2,3\}$ 
\beq
Z = \int {\mathcal D}(\pd){\mathcal D}(\psi){\mathcal D}(\Phi_a^{\prime})
\exp\Big\{-I(\pd,\psi)-\frac{1}{4 g_s} \int d^4 x_E(\Phi_a^{\prime}+2 g_s\pd\gamma_0\Gamma_a\psi)^2 \Big\}~,
\eeq
with $\Gamma_a = (1,i\gamma_5\vec{\tau})$. Then the action contains only
bilinear terms in the quark fields, so that they can be integrated out. 
After performing a shift of the auxiliary
fields, $\Phi_a = \Phi^{\prime}_a + (m_0,\vec{0})$, one finally arrives at the
bosonized action
\beq
 I(\Phi) =  -Tr\ln S^{-1} + \frac{1}{4 g_s}\int d^4x_E 
           (\Phi^2-2 m_0 \Phi_0+m_0^2)~,
\label{ba}
\eeq
where $S^{-1}$ is the Dirac operator
\beq
S^{-1} = \gamma_0 \partial_{\tau} - i\vec{\gamma}\cdot\vec{\nabla}
+\Gamma_a\Phi_a~.
\eeq
The symbol $Tr$ in \eq{ba} is to be
understood as a functional trace and a trace over internal degrees of freedom
like flavor, color and spin. $Tr \ln S^{-1}$ is the
quark-loop contribution. The imaginary part of
this term vanishes for the $SU(2)$ case and we can rewrite the action as:
\beq   
 I(\Phi) = -\frac{1}{2} Tr \ln S^{-1\dagger}S^{-1} + \frac{1}{4 g_s}\int d^4x_E (\Phi^2-2 m_0
 \Phi_0+m_0^2)~. 
\eeq

The effective action $\Gamma(\Phi)$ is defined as Legendre transform of 
the generating functional $W(j)$. 
Its stationary point $\ave{\Phi_a}$, i.e.
\beq
\frac{\delta\Gamma(\Phi)}{\delta\Phi_a}\Big|_{\Phi_a = \ave{\Phi_a}} = 0~,
\label{vev}
\eeq
represents the vacuum expectation values of the fields.

The quark condensate can be expressed via the expectation value of $\Phi_0$.
It is given as 
\beq
\qq = \frac{\partial W}{\partial m_0} = -\frac{1}{2 g_s} (\ave{\Phi_0}-m_0)~.
\label{qqea}
\eeq
Another important feature of the effective action is that the
inverse propagators of the fields (in our case the propagators for $\pi$- and
$\sigma$-mesons) can be generated in a symmetry conserving way 
by second-order derivatives
\beq
D^{-1}_{ab} =
\frac{\delta^2\Gamma(\Phi)}{\delta\Phi_a\delta\Phi_b}~. 
\label{mesonsea}
\eeq

To obtain an expression for the effective action the
path integral is evaluated using the saddle point approximation.
The lowest-order contribution to the effective action is 
\beq
\Gamma^{(0)}(\Phi) = I(\Phi)~.
\label{gamma0}
\eeq
This
corresponds to the mean-field (Hartree) approximation \cite{ripka}. The vacuum
expectation values of the fields in mean-field approximation coincide
with the stationary point of the action $I(\Phi)$. This is obvious if one
combines \eq{vev} and \eq{gamma0}.
Including quadratic mesonic fluctuations leads to the following expression for
the effective action \cite{ripka}: 
\beq
\Gamma(\Phi) \= I(\Phi) \+ \frac{1}{2}\,Tr \ln (\frac{\delta^2
                         I(\Phi)}{\delta\Phi_a\delta\Phi_b})~.
\label{onem}
\eeq
The second term in the above expression contains the mesonic fluctuations.
As discussed in Ref.~\cite{ripka} the method is only meaningful if
the second-order functional derivative which enters into this term is 
positive definite. Otherwise severe problems arise due to an ill-defined
logarithm, which would then be complex. We will come back to this point 
in \Sec{solution}.
 
Determining the stationary point of the effective action in \eq{onem} leads to
the following ``gap equation'' \cite{nikolov} 
\beq
\ave{\Phi_0}-m_0-\Sigma_H(\ave{\Phi_0})-\delta\tilde\Sigma(\ave{\Phi_0}) = 0~.
\label{qqneu}
\eeq
Here $\Sigma_H$ and $\delta\tilde\Sigma$ are the same functions we already 
defined in Eqs.~(\ref{gap}) and (\ref{deltatildesigma}) in the context of 
the Hartree- and the LSS gap equation. In fact, \eq{qqneu} is identical to
the LSS gap equation, \eq{localgap}, if we identify $\ave{\Phi_0}$ with
the LSS-constituent quark mass $m$. 

In the same way we exactly recover the meson structure of the LSS if
we evaluate \eq{mesonsea} at the stationary point.
This means, the ``local selfconsistent scheme'' which was constructed from 
a somewhat arbitrary starting point in \Sec{nonpert2} can be derived
in a systematic way in the effective action formalism. 
However, the interpretation is different: As emphasized in
Ref.~\cite{nikolov}, the solution of the gap equation is only the 
expectation value of the $\Phi_0$ field and does {\it not} correspond 
to the pole of the quark propagator. 
This becomes clear if we look at the quark condensate, 
which is given by \eq{qqea}. The r.h.s. of this equation is identical to 
\eq{qbqself2} and therefore different from \eq{qbqself1}, which was derived 
by taking the trace over what we called the ``quark propagator'' in 
\Sec{nonpert2}. 

Hence, within the effective action formalism, \eq{qbqself2} is the
correct expression for the quark condensate (in that approximation scheme), 
whereas the gap equation should not be interpreted as an equation for 
the corresponding inverse quark propagator. 
In the following, we will adopt this point of view. For simplicity, 
however, we will still call $m$ a ``constituent quark mass'' and
$(\psl - m)^{-1}$ a ``quark propagator'', although this is not
entirely correct.


\section{Consistency with chiral symmetry}
\label{secpion}
\setcounter{equation}{0}

By construction, the LSS is consistent with axial 
Ward-Takahashi identities and hence -- as discussed in \Sec{nonpert1} --
with the Goldstone theorem.
Since the mesonic polarization functions of the LSS
contain all diagrams up to next-to-leading order of the
$1/N_c$-expansion scheme and the various contributions to the pion mass
have to cancel order by order in the chiral limit, this implies that the 
$1/N_c$ scheme discussed in \Sec{model2} is also consistent with the 
Goldstone theorem. 

Nevertheless, for the numerical implementation it is instructive, to show
the consistency of the different schemes with chiral symmetry on a
less formal level. Since most of the integrals which have to be evaluated 
are divergent and must be regularized one has to ensure that the 
various symmetry relations are not destroyed by the regularization. 
To this end, it is important to know how these relations emerge in detail.
This will also enable us to perform approximations without violating
chiral symmetry. As we will see in \Sec{solution}, this is very important
for practical calculations within the LSS,
which cannot be applied as it stands.

For both, the $1/N_c$-expansion and the LSS,
we begin our discussion with the explicit proof of the Goldstone theorem.  
This was given first by Dmitra\v{s}inovi\'{c} et al.~\cite{dmitrasinovic}. 
After that we discuss the Gell-Mann--Oakes--Renner (GOR) relation. 
This is of particular interest in the context of the proper definition
of the quark condensate in the LSS 
(cf. Eqs.~(\ref{qbqself1}) and (\ref{qbqself2})).


\subsection{$1/N_c$-expansion}
\label{pionnc}
We begin with the $1/N_c$-expansion scheme. For the Goldstone theorem
one has to show that, in the chiral limit, the inverse pion propagator 
vanishes at zero momentum, 
\beq 
    2g_s\,{\tilde\Pi}_\pi(0) \= 1 \qquad {\rm for} \quad m_0 \= 0.
    \label{goldstone}
\eeq
As before we use the notation
${\tilde\Pi}^{ab}_\pi = \delta_{ab}{\tilde\Pi}_\pi$.  
The function ${\tilde\Pi_\pi}^{ab}$ has been defined in \eq{pol1}. It consists
of the RPA polarization loop  $\Pi_\pi^{ab}$ and the four $1/N_c$-correction 
diagrams $\delta\Pi_\pi^{(k)\,ab}$, $k = a,b,c,d$.
Restricting the calculation to the chiral limit and to zero momentum
simplifies the expressions considerably and Eq.~(\ref{goldstone}) can 
be proven analytically.

For the RPA loop one obtains 
\beq
    2g_s\,\Pi_\pi(0) = \frac{\Sigma_H}{m}  \;.
\label{pi0}
\eeq
This is the relation which guarantees the consistency of the 
Hartree + RPA scheme: In Hartree approximation we have 
$m = m_0 + \Sigma_H$ and hence \eq{goldstone}, is fulfilled by \eq{pi0}.
Since the gap equation is not changed in the perturbative $1/N_c$ expansion,
this remains true, if we include the next-to-leading order.  
Therefore we have to show that the contributions of the correction terms 
add up to zero:
\beq 
    \sum_{k=a,b,c,d}\;\delta\Pi_\pi^{(k)}(0) \= 0
    \qquad {\rm for} \quad m_0 \= 0.
    \label{deltapisum}
\eeq
The correction terms $\delta\Pi_\pi^{(k)}$ are defined in \eq{deltapi}.
Let us begin with diagram $\delta\Pi_\pi^{(a)}$. As mentioned above, 
we neglect the $\rho$ and $a_1$ subspace for intermediate mesons.
Then one can easily see that the external pion can only couple to
a $\pi\sigma$ intermediate state. Evaluating the trace in 
Eq.~(\ref{trianglevertex}) for zero external momentum one gets for
the corresponding triangle diagram:
\beq
  \Gamma_{\pi,\pi,\sigma}^{ab}(0,p) = -\delta_{ab}\ 4  N_c N_f\ 2 m\ I(p)~, 
  \label{gammapps}
\eeq
with $a$ and $b$ being isospin indices and the elementary integral
\beq
    I(p) = \intk \frac{1}{(k^2-m^2+i \eps)( (k+p)^2-m^2+i\eps)}\;.
\eeq
Inserting this into Eq.~(\ref{deltapi}) we find 
\beq
\delta\Pi_\pi^{(a)\,ab}(0) 
\= i \delta_{ab}\intp  (4 N_c N_f I(p))^2 4 m^2\ D_\sigma(p)\ D_\pi(p) \;.
\label{pisigtri}
\eeq
Now the essential step is to realize that the product of the RPA sigma- 
and pion propagators can be converted into a difference \cite{dmitrasinovic}, 
\beq
D_\sigma(p)\ D_\pi(p) =
  i \,\frac{D_\sigma(p)-D_\pi(p)} {4 N_c N_f\ 2 m^2\ I(p)}\;,
\label{pisig}
\eeq
to finally obtain
\bea
\delta\Pi_\pi^{(a)\,ab}(0) =-\delta_{ab}\;4 N_c N_f\intp
2 I(p)\Big\{ D_\sigma(p)-D_\pi(p)\Big\} \;. 
\label{pisigend}
\eea 
The next two diagrams can be evaluated straightforwardly.
One finds:
\bea
  \delta\Pi_\pi^{(b)\,ab}(0)&=&  -\delta_{ab}\;4 N_c N_f\intp\Big\{ 
  D_\sigma(p)\  \big(I(p)+I(0)-(p^2-4 m^2)\ K(p)\big) \nonumber \\
&&\hspace{3.5cm} +D_\pi(p)\ \big(3I(p)\hspace{0.2cm}+\;3I(0)\hspace{0.4cm}
-\;\;3p^2\ K(p)\big)\; \Big\} \;, 
\nonumber\\ 
  \delta\Pi_\pi^{(c)\,ab}(0)&=& -\delta_{ab} \;4 N_c N_f\intp I(p)\Big\{ 
  -D_\sigma(p) - D_\pi(p) \Big\} \;.
\label{pseudo} 
\eea
The elementary integral $K(p)$ is of the same type as the integral
$I(p)$ and is defined in App.~\ref{integrals}.

Finally we have to calculate $\delta\Pi_\pi^{(d)}(0)$.
According to Eq.~(\ref{deltapi}),  it can be written in the form 
\beq
    \delta\Pi_\pi^{(d)\,ab}(0) \= -i\Gamma^{ab}_{\pi,\pi,\sigma}(0,0) \,
    D_\sigma(0)\,\Delta  \;.
    \label{deltapid}
\eeq
For the constant $\Delta$, defined in \eq{Delta}, one obtains
\bea
 \Delta &=& 4 N_c N_f\ m \int\frac{d^4p}{(2\pi)^4}{\Big\{}
\ D_\sigma(p)\ (2\ I(p)+I(0)-(p^2-4 m^2)\ K(p)) \nonumber \\
&&\hspace{3.3cm} +  D_\pi(p)\  (\;3I(0)\;-\;3p^2\ K(p)\;) 
\hspace{2.5cm}{\Big\}}~. 
\label{tdself}
\eea
Evaluating $D_\sigma(0)$ in the chiral limit and comparing the result with
Eq.~(\ref{gammapps}) one finds that the product of the first two factors 
in Eq.~(\ref{deltapid}) is simply $\delta_{ab}/m$, i.e. one gets   
\beq
 \delta\Pi_\pi^{(d)\,ab}(0) = \delta_{ab} \,\frac{\Delta}{m}~.
\label{pid}
\eeq
With these results it can be easily checked that Eq.~(\ref{deltapisum}) 
indeed holds in this scheme. 

As already pointed out, most of the integrals we have to deal with 
are divergent and have to be regularized. 
Therefore one has to make sure that all
steps which lead to Eq.~(\ref{deltapisum}) remain valid in the 
regularized model.   
One important observation is that the cancellations occur already
on the level of the $p$-integrand, i.e. before performing the 
meson-loop integral. This means that there is no restriction on the
regularization of this loop. 
We also do not need to perform the various quark loop integrals explicitly
but we have to make use of several relations between them. For instance,
in order to arrive at Eq.~(\ref{pid}) we need the similar structure
of the quark triangle $\Gamma_{\pi,\pi,\sigma}(0,0)$ and the inverse RPA 
propagator $D_\sigma(0)^{-1}$. Therefore all quark loops, i.e. RPA 
polarizations, triangles and box diagrams should be consistently 
regularized within the same scheme, whereas the meson loops can be 
regularized independently.

Going away from the chiral limit the pion recieves a finite mass.
To lowest order in the current quark mass it is given by the 
Gell-Mann--Oakes--Renner (GOR) relation,
\beq
    m_\pi^2 \, f_\pi^2 \= -m_0 \,\ave{\pb\psi} \;.
\label{GOR}
\eeq
However, in the $1/N_c$-expansion scheme we cannot expect, that the GOR 
relation holds in this form. In \Sec{model} we have calculated the quark 
condensate in leading order and next-to-leading order in $1/N_c$.
Hence, to be consistent, we should also expand the l.h.s. of the
GOR relation up to next-to-leading order in $1/N_c$:  
\beq
m_{\pi}^{2(0)} f_{\pi}^{2(0)} \+ m_{\pi}^{2(0)}\delta f_{\pi}^2 
\+ \delta m_{\pi}^2 f_{\pi}^{2(0)}
= -m_0 \,  \Big(\ave{\pb\psi}^{(0)} + \delta\ave{\pb\psi}\Big) \;.
\label{gor1}
\eeq
Here, similar to the notations we already introduced for the 
quark condensate, $m_{\pi}^{2(0)}$ and $f_{\pi}^{2(0)}$ denote the
leading order and $\delta m_{\pi}^2$ and $\delta f_{\pi}^2$
the next-to-leading order contributions to the squared pion mass
and the squared pion decay constant, respectively. 
Since the GOR relation holds only in lowest order in $m_0$, Eq.~(\ref{gor1})
corresponds to a double expansion: $m_\pi^2$ has to be calculated in
linear order in $m_0$, $f_\pi^2$ and $\ave{\pb\psi}$ in the chiral limit.

The leading-order and next-to-leading-order expressions for the quark
condensate  
are given in Eqs.~(\ref{qbq0}) and (\ref{deltaqbqexp}).
The pion decay constant $f_{\pi}$ is calculated from the one-pion to vacuum
axial vector matrix element.
Basically this corresponds to evaluating the mesonic polarization diagrams,
Fig.~\ref{fig3}, coupled to an external axial current and to a pion.
This leads to expressions similar to Eqs.~(\ref{pol0}) and (\ref{deltapi}),
but with one external vertex equal to $\gamma^\mu \gamma_5 \frac{\tau^a}{2}$,
corresponding to the axial current, and the second external vertex
equal to $g_{\pi qq} i\gamma_5 \tau^b$, corresponding to the pion.
Here the $1/N_c$-corrected pion-quark coupling constant is 
defined as
\beq
    g_{\pi qq}^{-2} \= g_{\pi qq}^{-2(0)} \+ \dg \=
\frac{d{\tilde\Pi}_\pi(q)}{dq^2}|_{q^2 = m_\pi^2} \;,
\label{gpiqq}
\eeq
analogously to Eq.~(\ref{mesonmass0}). Now we take the divergence of the
axial current and then use the relation 
\beq
\gamma_5 \,\pslash \= 2 m \gamma_5 \,+ \gamma_5 \,S^{-1}(k+p) 
                   \,+ S^{-1}(k)\, \gamma_5
\label{axialward}
\eeq
to simplify the expressions~\cite{dmitrasinovic}. One finds:
\beq
    f_{\pi} \= g_{\pi qq} \; \Big(\; \frac{{\tilde \Pi}_\pi(q) - 
    {\tilde \Pi}_\pi(0)}{q^2}\,m \+ \frac{\Pi_\pi(q) - \Pi_\pi(0)}{q^2} 
    \, D_\sigma(0)\,\Delta\; \Big)\Big|_{q^2=m_\pi^2}\;. 
\label{fpi}
\eeq
In the chiral limit, $q^2 = m_\pi^2\rightarrow 0$,
Eqs.~(\ref{mesonmass0}) and (\ref{gpiqq}) can be employed to replace 
the difference ratios on the r.h.s. by pion-quark coupling 
constants.
When we square this result and only keep the leading order and the
next-to-leading order in $1/N_c$ we finally obtain:
\beq
f_{\pi}^{2(0)} \+ \delta f_{\pi}^2 \= 
m^2 \,g_{\pi qq}^{-2(0)} \+ \Big(\,m^2 \,\dg  \+
2m\,D_\sigma(0)\,\Delta\,g_{\pi qq}^{-2(0)}\,\Big)~.
\label{fpi2}
\eeq

For the pion mass we start from Eqs.~(\ref{dm1}) and (\ref{mesonmass1}) 
and expand the inverse pion propagator around $q^2=0$:
\beq
    1 \;-\; 2g_s\,{\tilde\Pi}_\pi(0) \;-\; 2g_s\, 
    \Big(\frac{d}{dq^2}\,{\tilde\Pi}_\pi(q)\Big)\Big|_{q^2=0}\,m_\pi^2 
    \+ {\cal O}(m_\pi^4) \= 0 \;.  
\eeq
To find $m_\pi^2$ in lowest non-vanishing order in $m_0$ we have to
expand $1-2g_s {\tilde\Pi}_\pi(0)$ up to linear order in $m_0$, while
the derivative has to be calculated in the chiral limit, where it can
be identified with the inverse squared pion-quark coupling constant,
Eq.~(\ref{gpiqq}). The result can be written in the form
\beq
    m_\pi^2 \= \frac{m_0}{m}\,\frac{g^2_{\pi qq}}{2g_s}\;
    \Big(\,1 \,-\,\frac{D_\sigma(0)\Delta}{m}\,\Big) 
    \+  {\cal O}(m_0^2) \;.
\label{mpi}   
\eeq
Finally one has to expand this equation in powers of $1/N_c$.
This amounts to expanding $g_{\pi qq}^2$, which is the only term in 
Eq.~(\ref{mpi}) which is not of a definite order in $1/N_c$.
One gets:
\beq
    m_\pi^{2(0)} \+ \delta m_\pi^2 \= 
    m_0 \,\frac{m}{2g_s}\,\frac{g^{2(0)}_{\pi qq}}{m^2} \;-\;
    m_0 \,\frac{m}{2g_s}\,\frac{g^{2(0)}_{\pi qq}}{m^2} 
    \Big(\,g_{\pi qq}^{2(0)}\,\dg 
    \,+\,\frac{D_\sigma(0)\Delta}{m}\,\Big) 
    \;.
\label{mpinc}   
\eeq
It can be seen immediately that the leading-order term is exactly equal to 
$-m_0 \ave{\pb\psi}^{(0)}/f_\pi^{2(0)}$, as required by the GOR relation. 
Moreover, combining Eqs.~(\ref{deltaqbqexp}), (\ref{fpi2}) and 
(\ref{mpinc}) one finds that the GOR relation in next-to-leading order, 
Eq.~(\ref{gor1}), holds in this scheme.

However, one should emphasize that this result is obtained by a strict 
$1/N_c$-expans\-ion of the various properties which enter into the GOR
relation and of the GOR relation itself. If one takes $f_\pi$ and $m_\pi$
as they result from Eqs.~(\ref{fpi}) and (\ref{mpi}) and inserts them 
into the l.h.s. of Eq.~(\ref{GOR}) one will in general find deviations from 
the r.h.s. which are due to higher-order terms in $1/N_c$.  
In this sense one can take the  violation of the GOR relation as a measure 
for the importance of these higher-order terms \cite{oertel}.


\subsection{Local selfconsistent scheme}
\label{piondstl}

The proof of the Goldstone theorem in the LSS is very similar to that
in the $1/N_c$-expansion scheme. Therefore, we can be brief, concentrating
on the steps which are different.

Again we have to show the validity of \eq{goldstone}. In the LSS
the function $\tilde\Pi_\pi$ is given by \eq{poltilde}, i.e. it differs 
from to the corresponding function in the $1/N_c$-expansion scheme
(\eq{pol1}) by the fact, that diagram $\delta \Pi_\pi^{(d)}$ is
(formally) missing. (As we already discussed it is implicitly contained
in the RPA diagram.) The other diagrams have the same structure as
before and we can largely use the results of the previous subsection.
However, we should keep in mind, that the constituent quark mass
is now given by the extended gap equation, \eq{localgap}.
Therefore, the r.h.s. of \eq{pi0} is different from unity in the chiral limit
and RPA pions are not massless. This has important consequences for the
practical calculations within this scheme, which will be discussed in greater 
detail in section~\ref{solution}.

Using Eqs.~(\ref{pisigend}), (\ref{pseudo}) and (\ref{tdself}) as well as
\eq{deltatildesigma} 
we get for the correction terms to the pion polarization function
\beq
    \sum_{k=a,b,c}\; \delta \Pi_\pi^{(k)}(0) 
    \= - \,\frac{\Delta}{m} \= \frac{\delta\tilde\Sigma}{2g_s m}
\;.
\eeq
Hence, together with the modified gap equation (\ref{localgap}) 
we find
\beq 
    2g_s\,{\tilde\Pi}_\pi(0) \= 1 \;-\; \frac{m_0}{m}
\eeq
in agreement with Eq.~(\ref{goldstone}).

The discussion concerning the regularization procedure can be repeated here. 
The structure of the proof again leads to the conclusion that we have to
regularize the quark loops in the same way, whereas we have the freedom to
choose the regularization for the meson loops independently.

Another important observation is that we, in both schemes, do not need the
explicit form of the
RPA propagators. $D_{\sigma}(p)$ and $D_{\pi}(p)$ only need to fulfill
Eq.~(\ref{pisig}). Thus, approximations to the RPA propagators can be made as
long as Eq.~(\ref{pisig}) remains valid.

For a non-vanishing current quark mass the pion mass is given by the 
GOR relation (Eq.~(\ref{GOR})). 
To linear order in $m_0$ this relation holds exactly in the LSS, if 
we choose the appropriate definition of the quark condensate. 
This will be demonstrated in the following. 

For the pion decay constant $f_{\pi}$ we follow the same steps as in the
$1/N_c$-expansion scheme to arrive at the following expression:
\beq
    f_{\pi} \= g_{\pi qq} \; m\, \frac{{\tilde \Pi}_\pi(q) - 
    {\tilde \Pi}_\pi(0)}{q^2}\Big|_{q^2=m_\pi^2}\;. 
\label{fpidstl}
\eeq
Here the modified pion-quark coupling constant is defined as
\beq
    g_{\pi qq}^{-2} \=
\frac{d{\tilde\Pi}_\pi(q)}{dq^2}|_{q^2 = m_\pi^2} \;.
\label{gpiqqdstl}
\eeq
In the chiral limit, $m_\pi^2\rightarrow 0$,
the difference ratio on the r.h.s. of Eq.~(\ref{fpidstl})
can be replaced by the pion-quark coupling constant
(Eq.~(\ref{gpiqqdstl}). This leads to the Goldberger-Treiman relation
\beq
f_{\pi} g_{\pi qq} = m~.
\label{gtr}
\eeq
Following the analogous steps which led us to \eq{mpi}
we find for the pion mass
\beq
m_{\pi}^2 = \frac{m_0}{m} \frac{g_{\pi qq}^2}{2 g_s}
            \+  {\cal O}(m_0^2) \;.
\eeq
Multiplying this with $f_\pi^2$ as given by \eq{gtr} we get to linear order
in $m_0$:
\beq
    m_\pi^2\,f_\pi^2 \= m_0 \,\frac{\tilde\Sigma}{2g_s} \;.
\eeq
Obviously this is consistent with the GOR relation (\eq{GOR}) if the
quark condensate is given by \eq{qbqself2}, but not if is given by 
\eq{qbqself1}. In \Sec{EAF} we have seen that within the effective action
formalism the quark condensate is given by \eq{qbqself2}. 
Therefore at this point we clearly see that the interpretation of $m$ as a constituent
quark mass, which would mean that we have to calculate the quark condensate
according to \eq{qbqself1}, leads to a contradiction with the GOR
relation. 
Therefore, in the numerical part, we will calculate the quark condensate 
according to \eq{qbqself2}.


\section{Numerical results at zero temperature}
\label{numerics}
\setcounter{equation}{0}

In this section we present our numerical results at zero temperature. 
We begin with a brief description of the regularization scheme and 
then discuss peculiarities related to the solution of the gap 
equation in the LSS. 
After that we study the influence of mesonic fluctuations on quantities
in the pion sector, thereby focusing on possible instabilities. 
Finally we perform a refit of these quantities within the
{\nce} and the LSS and apply the model to observables in the 
$\rho$-meson sector.


\subsection{Regularization}
\label{regularization}

Before we begin with the explicit calculation we have to fix our
regularization scheme.
As discussed in Sec.~\ref{secpion}, all quark loops, 
i.e. the RPA polarization diagrams, the quark triangles and the quark 
box diagrams must be regularized in the same way in order to preserve 
chiral symmetry.
We use Pauli-Villars-regularization with two regulators, i.e. we replace
\beq
    \intk f(k;m) \;\rrr\; \intk \sum_{j=0}^2 c_j\,f(k;\mu_j)~,
    \label{pv}
\eeq
with
\beq
    \mu_j^2 \= m^2 \+ j\,\Lambda_q^2~;  \qquad
    c_0 = 1, \quad c_1 = -2, \quad c_2 = 1~.
\eeq
Here $\Lambda_q$ is a cutoff parameter. 

The regularization of the meson loop
(integration over $d^4p$ in Eq.~(\ref{deltapi})) is not constrained by 
chiral symmetry and independent from the quark loop regularization. 
For practical reasons we choose a three-dimensional cutoff $\Lambda_M$ 
in momentum space. In order to obtain a well-defined result we work in the 
rest frame of the ``improved''  meson.
The same regularization scheme was already used
in Refs.~\cite{oertel,OBW}. 


\subsection{Solution of the gap equation in the LSS}
\label{solution}

In contrast to the $1/N_c$-expansion scheme, where all diagrams are
constructed from ``Hartree'' quarks, the LSS is based on the extended 
gap equation, \eq{localgap}.
In \Sec{nonpert2} this equation was the starting point to find a
consistent set of diagrams for the description of mesons. 
In fact, in \Sec{piondstl} we have shown, that various symmetry relations,
namely the Goldstone theorem, the Goldberger Treiman relation and
the GOR relation hold in this scheme. 
It is not surprising, that the structure of the extended gap equation
was needed to prove these relations. So far, all this has been done
on a rather formal level. This section is now devoted to the explicit 
solution of the modified gap equation in the LSS.
We will see, that this cannot be done in a straightforward
manner and we are forced to a slight modification of the scheme. 

\begin{figure}[b!]
\parbox{6cm}{
     \epsfig{file=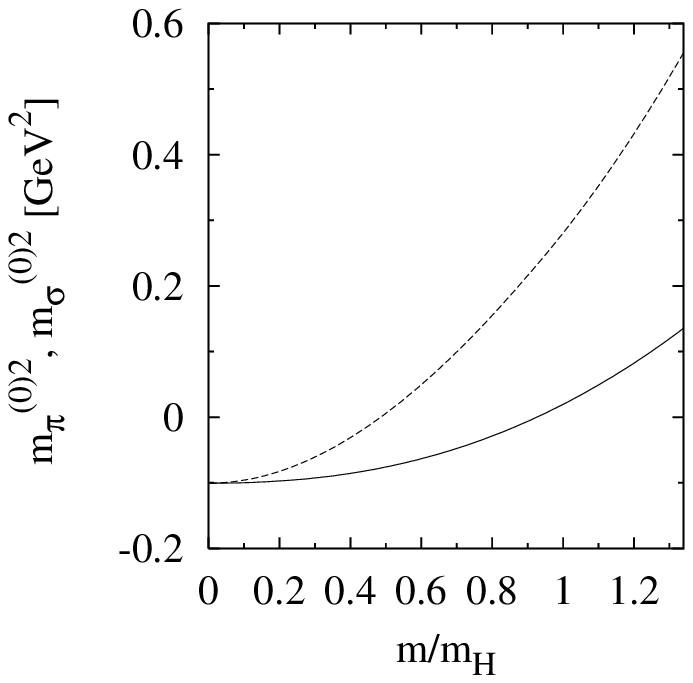,
     height=6cm, width=7.cm}\quad}
\hspace{2cm}
\parbox{6cm}{
     \epsfig{file=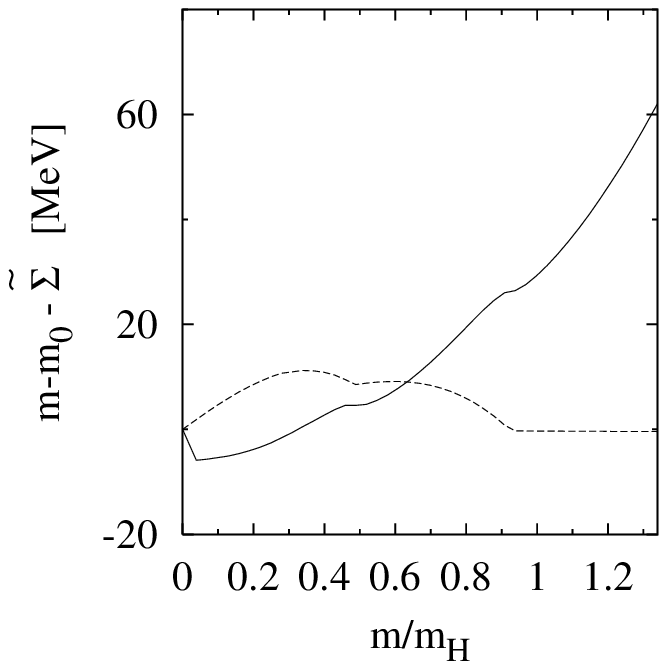,
     height=6cm, width=7cm}\quad}
\hspace{1cm}
\caption{\it (Left) Squared pole masses of the pion (solid) and the
         $\sigma$-meson (dashed) in RPA as functions of a trial 
         constituent quark mass $m'$ in units of the Hartree quark mass.
         (Right) Difference $m' - m_0 - \tilde\Sigma(m')$
         between the l.h.s. and the r.h.s. of the LSS gap equation, 
         \eq{localgap}, 
         as a function of the trial constituent quark mass $m'$.
         The real part is denoted by the solid line, the imaginary part
         by the dashed line.}
\label{figqm} 
\end{figure}

In addition to the Hartree term $\Sigma_H$,  \eq{localgap} contains
the term $\delta\tilde\Sigma$, which is a quark loop, dressed by
RPA mesons (see \fig{fig8}). 
As already pointed out, these RPA mesons consist of quarks with
the selfconsistent mass $m$, which is in general different from the
``Hartree'' mass $m_H$. Hence, the masses of these mesons are
also different from the meson masses in the Hartree + RPA scheme. 
On the l.h.s. of \fig{figqm} we have plotted the squared masses 
$m_M^{(0)\,2}$ of the RPA pion (solid) and the RPA $\sigma$-meson (dashed)
as functions of a trial constituent quark mass $m'$. An important observation 
is that the pion becomes tachyonic, i.e. $m_{\pi}^{(0)2}$ becomes negative, 
for quark masses smaller than the Hartree quark mass. Strictly speaking this 
is only the case in the chiral limit, whereas for nonvanishing current quark 
masses, $m_{\pi}^{(0)2}$ becomes negative slightly below the Hartree quark 
mass. A similar observation can be made for $m_{\sigma}^{(0)2}$, but 
only for $m'$ much smaller than the Hartree mass.
This observation of tachyonic RPA mesons is related to the point discussed in
\Sec{EAF}, that the meson-loop term in the effective action (second term of
\eq{onem}) is no longer positive definite.  

Tachyonic RPA mesons lead to a complex correction to the quark self-energy.
Therefore the solution of the extended gap equation can only be real if it 
is larger than the Hartree mass. Otherwise it must be complex.
To investigate this point we plot the difference between the l.h.s. and 
the r.h.s. of \eq{localgap} as a function of the (real) trial quark mass
$m'$. This is shown in the right panel of \fig{figqm}.
The solid line denotes the real part, the dashed line the imaginary part
of $m' - m_0 - \tilde\Sigma(m')$. Obviously, below the Hartree quark 
mass, the self-energy indeed gets complex. Moreover, we see that there is 
no solution of the gap equation for real constituent quark masses. 
Hence, in principle, one should search for solutions of the gap equation 
in the complex plane. However, this would mean that the RPA mesons
would also consist of quarks with complex masses. In this case, e.g.
a reasonable description of the $\rho$-meson would be completely impossible,
because its properties are mainly determined by intermediate pions.

Therefore, we prefer to perform an approximation, which was introduced
in Ref.~\cite{nikolov}. As discussed in \Sec{piondstl} the symmetry 
properties of the LSS are not affected by approximations to the RPA 
meson propagators which preserve the validity of Eq.~(\ref{pisig}). 
The authors of Ref.~\cite{nikolov} simply replace the RPA pion propagator 
in the extended gap equation
\beq
    D_{\pi}(p) \= -2g_s\; 
    \Big [\; 1\;-\;2 i g_s\;4 N_c N_f \intk \frac{1}{k^2 - m^2 + i\varepsilon}
    \+ 2i g_s (2 N_c N_f)\, p^2 \,I(p) \,\Big ]^{-1}
\eeq
by 
\beq
    D_{\pi}(p) \= -2g_s\; 
    \Big [\,\frac{m_0}{m} 
    \+ 2i g_s\;2 N_c N_f, p^2 \,I(p) \, \Big ]^{-1}
\label{dpirepl}
\eeq
and analogously for the $\sigma$-propagator.
The same replacements are performed for the RPA meson propagators in 
the correction terms $\delta\Pi_M^{(k)}$ to the mesonic polarization
diagrams. The RPA contribution $\Pi_M$ itself, however, is not changed.
In this way the solution of the gap equation and the masses of the 
intermediate mesons remain real. Moreover, in the chiral limit the
intermediate pions are massless, as one can immediately see from
\eq{dpirepl}. 

The above replacements would be exact in the Hartree approximation
(cf. \eq{gapex}). 
The authors of Ref.~\cite{nikolov} argue that the correction terms are 
suppressed because they are of higher orders of $1/N_c$ (beyond 
next-to-leading order). In the LSS, this is a questionable argument 
because the selfconsistent solution of the gap equation mixes all 
orders of $1/N_c$ anyway. Nevertheless this approximation preserves
the validity of the various symmetry relations we have checked in 
\Sec{piondstl}.

In the following we will call this scheme, including the above replacements,
the ``local selfconsistent scheme'' although it is strictly speaking only 
an approximation to the LSS as it was originally introduced in \Sec{nonpert2}.


\subsection{Meson-loop effects on quantities in the pion sector}
\label{secpi1}

In this subsection we want to study the influence of mesonic fluctuations
on the quark condensate, the pion mass and the pion decay constant, both
within the {\nce} and within the LSS. 
Since the strength of the fluctuations is controlled by the meson cutoff
$\Lambda_M$, we first keep all other parameters fixed and investigate how
the above quantities change, when $\Lambda_M$ is varied.
For the {\nce} this has been done in more detail in Ref.~\cite{oertel}.
Later, in the next subsection, we will perform a refit of the parameters to 
reproduce the empirical values of $\qq$, $m_{\pi}$ and $f_{\pi}$. 

Our starting point is the Hartree + RPA scheme, which corresponds to
$\Lambda_M = 0$. Here we obtain a reasonable fit 
($\ave{\pb\psi}^{(0)}$~=~-2~(241.1~MeV)$^3$, $m_\pi^{(0)}$~=~140.0~MeV 
and $f_\pi^{(0)}$~=~93.6~MeV) with the parameters $\Lambda_q$~=~800~MeV, 
$g_s\Lambda_q^2$~=~2.90 and $m_0$~=~6.13 MeV.
These parameters correspond to a relatively small ``Hartree'' constituent 
quark mass of 260~MeV.

\begin{figure}[t] 
\parbox{16cm}{
  \hspace{-1.cm}
              \epsfig{file=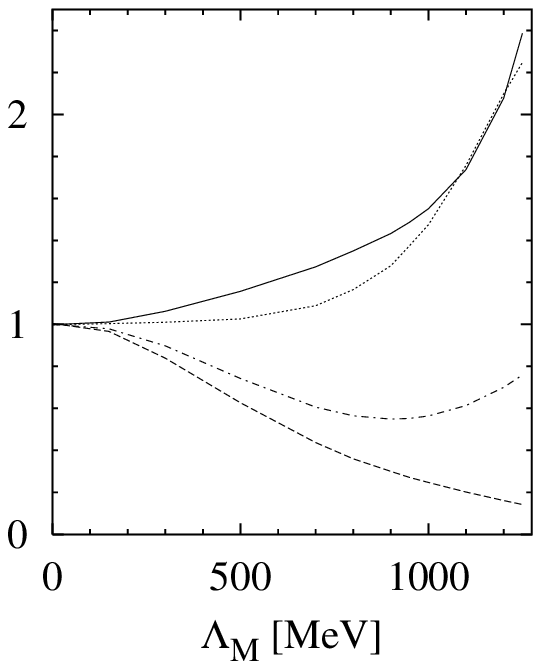,height=6cm,width=7cm}
              \hspace{2.cm}
              \epsfig{file=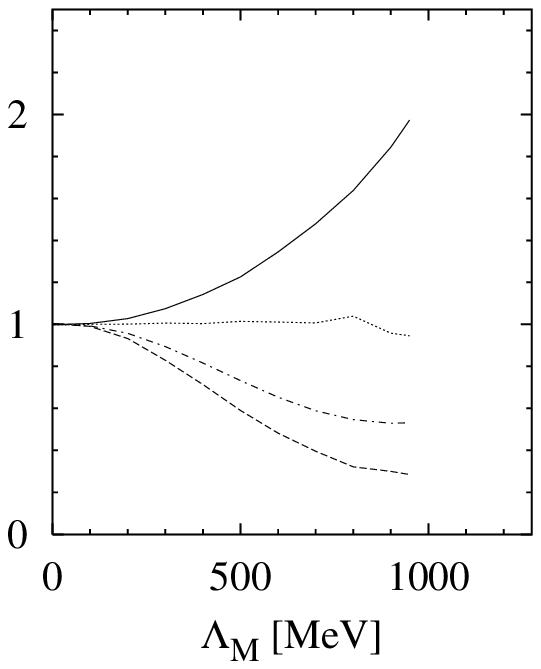,height=6cm,width=7cm}}
\caption{\it The ratios $m_\pi^2/{m_\pi^2}^{(0)}$ (solid),
         $f_\pi^2/{f_\pi^2}^{(0)}$ (dashed), 
         $\ave{\pb\psi}/\ave{\pb\psi}^{(0)}$ (dashed-dotted)
         and the combination
         $-m_0\ave{\pb\psi}/m_{\pi}^2f_{\pi}^2$ (dotted)  
         as a function of the meson loop cutoff $\Lambda_M$. 
         Left: \nce, Right: LSS}
\label{figpion}
\end{figure}

Now we turn on the mesonic fluctuations by taking a non-zero meson cutoff 
$\Lambda_M$. All other parameters are kept constant at the values given above.
The resulting behavior of $m_\pi^2$, $f_\pi^2$ and the quark condensate 
as a function of $\Lambda_M$ is displayed in Fig.~\ref{figpion}.
The left panel corresponds to the \nce, the right panel to the LSS.
As one can see, in both schemes the mesonic fluctuations lead to a reduction 
of $f_\pi$ (dashed lines) while $m_\pi$ (solid) is increased.
At smaller values of $\Lambda_M$ the absolute value of the quark condensate 
decreases but goes up again for $\Lambda_M  \gsim$~900~MeV. 
This is also an effect which is found in both schemes. 

In the Hartree + RPA scheme the quantities 
${m_\pi^2}^{(0)}$, ${f_\pi^2}^{(0)}$ 
and $\ave{\pb\psi}^{(0)}$, are in almost perfect agreement with the GOR 
relation, Eq.~(\ref{GOR}). 
As discussed in \Sec{pionnc}, the {\nce} is consistent with the GOR relation
up to next-to-leading order in $1/N_c$, but the relation is violated by
higher-order terms. We therefore expect a less perfect agreement in this
scheme, becoming worse with increasing values of $\Lambda_M$. In the LSS, 
on the other hand, the quantities should be in good agreement with the GOR 
(see \Sec{piondstl}).

These expectations are more or less confirmed by the results.
In Fig.~\ref{figpion}, the ratio of the  r.h.s. and the l.h.s. of
Eq.~(\ref{GOR}) is displayed by the dotted lines. In the $1/N_c$-expansion
scheme (left panel) the relation holds within 30\% for $\Lambda_M\leq$~900~MeV.
However, when the meson cutoff is further increased the deviation
grows rapidly. This indicates that higher-order corrections in $1/N_c$
become important in this regime and this perturbative scheme should no longer 
be trusted. In the LSS the agreement with the GOR is almost perfect.

In \fig{figpion} the various curves are only shown up to $\Lambda_M=1250$~MeV 
for the {\nce} and $\Lambda_M = 950$~MeV for the LSS. 
For larger values of $\Lambda_M$ a second, 
unphysical, pole with a residue of the ``wrong'' sign, emerges in the pion
propagator.
This would correspond to an imaginary pion-quark coupling constant and hence
an imaginary pion decay constant. 
Upon further increasing $\Lambda_M$ the two poles merge and finally disappear 
from the positive real axis.

For the $1/N_c$-expansion scheme this has been discussed in more details in 
Ref.~\cite{oertel}.
In that reference we suggested that the instabilities of the pion propagator 
might indicate an instability of the underlying ground state against mesonic
fluctuations.
In fact, it has been claimed by Kleinert and Van den Bossche \cite{kleinert}
that there is no spontaneous chiral symmetry breaking in the NJL model as
a consequence of strong mesonic fluctuations. Although this cannot be true in
general if the strength of the mesonic fluctuations is controlled by an 
independent cutoff parameter $\Lambda_M$ \cite{oertel}, this phenomenon might 
very well occur for large values of $\Lambda_M$. In other words: There could
be some kind of ``chiral symmetry restoration'' at a certain value of the
parameter $\Lambda_M$. 

Clearly, this could not be studied within the {\nce} where the mesonic 
fluctuations are built perturbatively on the Hartree ground state.
In the LSS, however, where we encounter the same type of instabilities in
the pion propagator, this question can be investigated more closely. 
To that end we consider the effective action \eq{onem}, which describes the 
energy density of the system. It is explicitly given by  
\bea
\Gamma(m^{\prime}) &=& -4i N_c N_f \intp \ln(\frac{m^{\prime2}-p^2}{m_0^2-p^2}) 
+ \frac{(m^{\prime}-m_0)^2}{4 g_s} \nonumber\\ &&-
\frac{i}{2}\intp\{ \ln (1- 2 g_s\Pi_{\sigma}(p))+3 \ln (1- 2
g_s\Pi_{\pi}(p))\} + const.~.
\label{tp}
\eea 
The irrelevant constant can be chosen in such a way that $\Gamma(0) = 0$.
The positions of the extrema of $\Gamma(m^{\prime})$ correspond to the solutions of the gap 
equation~(\ref{localgap}). In particular, the vacuum expectation value 
$m$ is given by the value of $m^{\prime}$ at the absolute minimum of 
$\Gamma$. Note that, according to \eq{qbqself2}, $m$ is proportional to the 
quark condensate, i.e. to the order parameter of chiral symmetry breaking.
Hence, for a given value of $\Lambda_M$, chiral symmetry is spontaneously
broken, if the absolute minimum of $\Gamma$ is located at a non-zero value 
of $m'$ and it is unbroken (``restored'') otherwise. 

\begin{figure}[t]
\begin{center} 
\parbox{8cm}{
              \epsfig{file=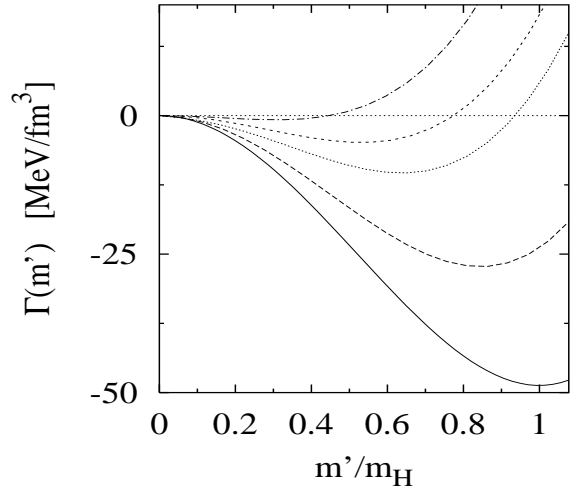,height=6.86cm,width=8cm}}\end{center}
\caption{\it  Effective action $\Gamma(m^{\prime})$ as a function of 
              $m^{\prime}/m_H$ in the LSS for different values of
              the meson cutoff $\Lambda_M$: 0~MeV (solid), 300~MeV
              (long-dashed), 500~MeV (dotted), 900~MeV (dashed-dotted) and
              1200 MeV (short-dashed). }
\label{figtp}
\end{figure}

We perform the calculations in the chiral limit.\footnote{To be precise,
we proceed as follows: Starting from the parameters given above, we keep
the Hartree constituent quark mass, $m_H =$~260~MeV, fixed, while $m_0$ 
is reduced from 6.1~MeV to zero.
Therefore the coupling constant is slightly enhanced from 
$g_s\Lambda_q^2 = 2.90$ to $g_s\Lambda_q^2= 2.96$.}
Our results for $\Gamma(m^{\prime})$ as a function of $m^{\prime}/m_H$ 
for different values of $\Lambda_M$ are displayed in \fig{figtp}.
For $\Lambda_M = 0$ we find of course the minimum at 
$m^{\prime}=m_H=260$~MeV, while there is a maximum at $m^{\prime}=0$.
If there was indeed a ``phase transition'' due to mesonic fluctuations,
this maximum should eventually convert to a minimum when $\Lambda_M$ is
increased.
In fact, for $\Lambda_M \lsim$~900~MeV the results seem to point in this
direction: In this regime the constituent quark mass $m$ is reduced
to about 30\% of the Hartree mass. At the same time the ``bag constant'' 
$B = \Gamma(0) - \Gamma(m)$ decreases from 48.7~MeV/fm$^3$ at $\Lambda_M$~=~0
to 0.8~MeV/fm$^3$ at $\Lambda_M$~=~900~MeV. 
However, upon further increasing $\Lambda_M$, both $m$ and $B$ go up
again. In particular, the point $\Gamma(0)$ always remains a local maximum:
In the LSS we do not observe a ``phase transition'' due to strong mesonic 
fluctuations. 

Here we should remark, that, according to the conjecture by Kleinert and
Van den Bossche \cite{kleinert}, the mesonic fluctuations do not restore the
trivial vacuum in the NJL model, but lead to a so-called pseudo-gap phase.
(See also Ref.~\cite{babaev} for a critical discussion of this article.)
In that phase the quarks still have a non-vanishing constituent mass, if
the latter is identified with the vacuum expectation value of the 
{\it modulus} of the scalar field $\Phi$ (cf. \Sec{EAF}). Nevertheless
chiral symmetry is not broken as the {\it phase} of the $\Phi$ field is 
strongly fluctuating. An analogous phenomenon is well known from 
strong-coupling superconductors above $T_c$ \cite{nozieres,sademelo}, 
where Cooper pairs are formed, but do not condense.
Obviously our above investigations, which focused on a the change
of $m$ assuming a uniform phase factor, cannot exclude a transition
into a phase of this type. Here more refined investigations are needed
to give a conclusive answer.
  
Another type of vacuum instability which is caused by unphysical poles
of the RPA meson propagators has recently been discussed by 
Ripka~\cite{ripkapaper}. Here ``unphysical'' means that these poles
are located in regions of the complex plane, where they are forbidden
by microcausality. Ripka stated, that they are induced by the
regulator scheme, in his case a 4-momentum cutoff or a Gaussian form factor.
In fact, the RPA meson propagators have this unphysical feature for most 
of the known regulator schemes, such as proper-time regularization, 
subtracted dispersion relations, dimensional regularization or, as 
mentioned above, a 4-momentum cutoff. A 3-momentum cutoff and 
Pauli-Villars-regularization in the form we use it (cf. 
appendix~\ref{correlators}) are exceptions.
On the other hand, due to Pauli-Villars regulators, the imaginary part
of the quark loops can have the wrong sign in some kinematical regions
and we cannot rule out that the instabilities we find for the pion 
propagator are related to this. 
This supposition is corroborated by the fact that these instabilities 
could be traced back to the imaginary part of the diagram $\delta\Pi^{(b)}$ 
(see \fig{fig4}) which has the ``wrong'' sign and which becomes large at
large values of $\Lambda_M$~\cite{oertel}. 
Further investigations are needed, however, to clarify this point.

Recently a second (unphysical) pole in the pion propagator has also been 
found in a non-local generalization of the NJL model~\cite{plant}.
The calculations indicate that these instabilities could probably be removed 
by including vector and axial vector intermediate states. This point is 
certainly worth a closer examination. 
In any case, at least in the {\nce} we found~\cite{OBW} that with a 
reasonable fit of all parameters we are far away from the region where these
instabilities occur. We will come back to this point in \Sec{secrho}.


\subsection{Parameter fit in the pion sector}
\label{secpi2}

In the previous subsection we did not change the parameters which were 
determined in the Hartree + RPA scheme by fitting $f_\pi^{(0)}$, $m_\pi^{(0)}$ 
and $\ave{\pb\psi}^{(0)}$.
Of course, if one wants to apply the model to describe physical processes 
a refit of these observables should be performed including the 
mesonic fluctuations.
In Ref.~\cite{OBW} this was already done for the {\nce} and we will now
try to perform an analogous fit within the LSS. 
Of course, by fitting the above three observables, we cannot conclusively
fix the five parameters of our model, $g_s$, $g_v$, $\Lambda_q$, 
$\Lambda_M$ and $m_0$. 
Therefore we try to proceed in a similar way as in Ref.~\cite{OBW}:
For various values of $\Lambda_M$ we fix the scalar coupling constant $g_s$, 
the current quark mass $m_0$ and the quark-loop cutoff $\Lambda_q$ to 
reproduce the empirical values of the pion mass, $f_{\pi}$ and $\qq$. 
Then, in the next subsection, we will try to fix the two remaining 
parameters, i.e. the vector coupling constant $g_v$ and the meson cutoff 
$\Lambda_M$, by fitting the pion electromagnetic form factor in the 
time-like region, which is related, via vector meson dominance, 
to the $\rho$-meson propagator. 
Roughly speaking, this amounts to fitting the $\rho$-meson mass and its width. 
Since in our model the latter is due to intermediate RPA pions,
we decided to fix the empirical value of $m_{\pi}^{(0)}$, not $m_{\pi}$,
in order to get the correct threshold behavior. In Ref.~\cite{OBW} we
found for the {\nce} that the deviation is about $10\%$.
As we will see below, in the LSS the difference is somewhat larger. 

Of course, the $\rho$-meson can only be described reasonably if the 
unphysical $q\bar{q}$-threshold lies well above the peak in the $\rho$-meson 
spectral function. i.e. the constituent quark mass $m$ should be larger 
than about 400 MeV. For that reason we try to increase the constituent 
quark mass as much as possible. Here we have some freedom as the
empirical value of the quark condensate is not known very precisely.
(Its absolute value is probably less than 2(260~MeV)$^3$, which
corresponds roughly to the upper limit extracted in Ref.~\cite{dosch} 
from sum rules at a renormalization scale of 1~GeV. Recent lattice 
results give $\qq$~=~{\mbox -2($(231 \pm 4 \pm 8 \pm 6)$~MeV)$^3$} \cite{giusti}.)
On the other hand, since the correction term $\delta\tilde\Sigma$ in
the LSS gap equation, \eq{localgap}, contributes negatively to $m$, 
it is much more difficult in the LSS to obtain sufficiently large quark masses
than in the \nce.   

\begin{table}[h!]
\begin{center}
\begin{tabular}{|c|c|c|c|c|c|}
\hline
$\Lambda_M$~/~MeV   &   0.  & 300.  & 500.  & 600.  & 700.  \\ \hline
$\Lambda_q$~/~MeV   & 800.  & 800.  & 800.  & 820.  & 852.  \\ \hline
$m_0$~/~MeV         & 6.13  & 6.40  & 6.77  & 6.70  & 6.54  \\ \hline
$g_s\Lambda_q^2$    & 2.90  & 3.07  & 3.49  & 3.70  & 4.16  \\ \hline
$m$~/~MeV           & 260.  & 304.  & 396.  & 446.  & 550.  \\ \hline
$m_\pi^{(0)}$~/~MeV & 140.0 & 140.0 & 140.0 & 140.0 & 140.0 \\ \hline
$m_\pi$~/~MeV       & 140.0 & 143.8 & 149.6 & 153.2 & 158.1 \\ \hline
$f_\pi^{(0)}$~/~MeV &  93.6 & 100.6 & 111.1 & 117.0 & 126.0 \\ \hline
$f_\pi$~/~MeV       &  93.6 &  93.1 &  93.0 &  93.1 & 93.4  \\ \hline
$\qq^{(0)}$~/~MeV$^3$ & -2(241.1)$^3$ & -2(249.3)$^3$ & -2(261.2)$^3$ 
                      & -2(271.3)$^3$ & -2(287.2)$^3$ \\ \hline
$\qq$~/~MeV$^3$       & -2(241.1)$^3$ & -2(241.7)$^3$ & -2(244.1)$^3$ 
                      & -2(249.5)$^3$ & -2(261.4)$^3$ \\ \hline
-$m_0 \qq / m_\pi^2 f_\pi^2$ & 1.001 & 1.007 & 1.018 & 1.023 & 1.072
\\ \hline
\end{tabular}
\end{center}
\caption{{\it The model parameters ($\Lambda_M$, $\Lambda_q$, $m_0$ and 
$g_s$) and the resulting values of $m_\pi$, $f_\pi$ and
$\qq$ (together with the corresponding leading-order quantities), 
the constituent quark mass $m$ in the \nce. The ratio $-m_0 \qq / m_\pi^2 f_\pi^2$,
is also given.
}}
\label{tablence}
\end{table}  
\begin{table}[h!]
\begin{center}
\begin{tabular}{|c|c|c|c|c|c|}
\hline
$\Lambda_M$~/~MeV   &   0.  & 300.  & 500.  & 600.  & 700.  \\ \hline
$\Lambda_q$~/~MeV   & 800.  & 800.  & 810.  &820. & 835.  \\ \hline
$m_0$~/~MeV         & 6.13  & 6.47 & 7.02  & 7.30 & 7.90  \\ \hline
$g_s\Lambda_q^2$    & 2.90  & 3.08 & 3.44  & 3.71 & 4.52  \\ \hline
$m_H$~/~MeV         & 260.  & 305. & 390.  & 450. & 600.  \\ \hline
$m$~/~MeV           & 260.  & 278.2 & 320.0 & 355.7& 468.4  \\ \hline
$m_\pi^{(0)}$~/~MeV & 140.0 & 139.9 & 140.0 & 139.7& 140.0 \\ \hline
$m_\pi$~/~MeV       & 140.0 & 145.1 & 156.3 & 164.5& 182.7 \\ \hline
$f_\pi^{(0)}$~/~MeV &  93.6 & 96.7 & 103.6 & 108.4& 120.0\\ \hline
$f_\pi$~/~MeV       &  93.6 & 93.2  &  92.9 & 92.9 & 92.8  \\ \hline
$\qq'$~/~MeV$^3$    & -2(241.1)$^3$ & -2(244.7)$^3$ & -2(254.3)$^3$  
                       &-2(261.9)$^3$ &  -2(277.3)$^3$ \\ \hline
$\qq$~/~MeV$^3$     & -2(241.1)$^3$ & -2(241.7)$^3$ & -2(246.2)$^3$ 
                       &- 2(250.8)$^3$ & -2(260.9)$^3$\\ \hline
-$m_0 \qq / m_\pi^2 f_\pi^2$ & 1.001 & 1.001& 1.006 &1.01& 1.02
\\ \hline
\end{tabular}
\end{center}
\caption{\it The same as in Table~\ref{tablence} for the LSS.
The quantity $\qq'$ denotes the quark condensate calculated according to
\eq{qbqself1}.}
\label{tablelss}
\end{table}  

Our results for the LSS are given in Table~\ref{tablelss}.
For comparison we also summarize the results obtained in Ref.~\cite{OBW}
within the {\nce} (Table~\ref{tablence}). In both tables we list 
five parameter sets (corresponding to five different meson cutoffs 
$\Lambda_M$), together with the constituent quark mass $m$, 
the values of $m_{\pi}, f_{\pi}$ and $\qq$ and the corresponding RPA 
quantities. In the LSS the ``RPA quantities'' are calculated with the 
constituent quark mass $m$ in order to represent the properties of the 
intermediate pion states. For completeness we also give the value of the
Hartree mass $m_H$ in Table~\ref{tablelss} and the value of the quark
condensate according to \eq{qbqself1}. We also show the ratio 
$-m_0 \qq / m_\pi^2 f_\pi^2$, which would be equal to 1 if the GOR relation 
was exactly fulfilled. Note that the deviations in the {\nce} are
less than 10\% (for $\Lambda_M \leq$~600~MeV even less than 3\%),
indicating that higher-order corrections in $1/N_c$ are small. 
In the LSS the deviations are considerably smaller, as already discussed
in the previous subsection.

In both schemes we find that the constituent quark mass increases with an
increasing meson cutoff $\Lambda_M$. In the {\nce} for 
$\Lambda_M \geq 500$~MeV
the value of $m$ is large enough to shift the $q\bar{q}$-threshold above the
$\rho$-meson peak. Besides, it turns out that we can only stay below the limit
of $-2 (260 {\mathrm MeV})^3$ for the quark condensate and simultaneously
reproduce the empirical value of $f_{\pi}$ if the cutoff is not too large
($\Lambda_M \lsim 700$ MeV). 
In the LSS the region of values for $\Lambda_M$
where on one hand the constituent quark mass is large enough and on the other
hand the quark condensate stays below the limit is much more narrow. This can be seen from the
values listed in Table~\ref{tablelss}. For a meson cutoff of
$\Lambda_M=600$~MeV $m$ is still too small and for $\Lambda_M=700$~MeV the
quark condensate lies already slightly above the limit. The reason for this is
obvious: In the LSS $m$ and the quark condensate are directly related
by \eq{qbqself2} and therefore the mesonic fluctuations which lower the quark
condensate also decrease the constituent quark mass. In the {\nce},
on the contrary, the meson loop effects only contribute to the quark condensate
and lower its value whereas $m$ is kept fixed at its Hartree value. 

\subsection{Description of the $\rho$-meson}
\label{secrho}

As already pointed out, the parameter fit in the pion sector was not complete. 
It is clear, e.g., that the meson-loop cutoff $\Lambda_M$ cannot be determined
just by fitting the pion mass, $f_\pi$ and $\qq$, since these observables
can already be reproduced in the Hartree + RPA scheme, i.e. without any
meson-loop effects. We only found an upper limit of 
$\Lambda_M \sim$~700~MeV in both schemes (see Tables~\ref{tablence} and
\ref{tablelss}). 
In Ref.~\cite{OBW} we therefore fixed the remaining parameters
$g_V$ and $\Lambda_M$ for the {\nce} in the $\rho$-meson sector.
In this subsection we want to give a short summary of these results
and then try to perform a similar fit for the LSS. 

According to Eqs.~(\ref{pol1}) and (\ref{poltilde}), the polarization 
function of the $\rho$-meson reads:
\beq
    {\tilde \Pi}_\rho^{\mu\nu, ab}(q) \= \Pi_\rho^{\mu\nu, ab}(q)
\+ \sum_{k}\; \delta \Pi_\rho^{(k)\;\mu\nu, ab}(q)
\;.
\label{rhopol1}
\eeq
Here $k$ runs over $\{a,b,c,d\}$ in the {\nce} and only over $\{a,b,c\}$
in the LSS. 
Because of vector current conservation, the polarization function 
has to be transverse, i.e.
\beq
    q_\mu\,{\tilde \Pi}_\rho^{\mu\nu, ab}(q) \=
    q_\nu\,{\tilde \Pi}_\rho^{\mu\nu, ab}(q) \= 0 \;.
\label{trans}
\eeq
With the help of Ward identities it can be shown that these relations hold
in both schemes, if we assume that the regularization preserves this property.
This is the case for the Pauli-Villars regularization scheme, which was 
employed to regularize the RPA part ${\Pi}_\rho$. 
Together with Lorentz covariance this leads to Eq.~(\ref{pirho}) for
the tensor structure of ${\Pi}_\rho$.
On the other hand, since we use a three-dimensional sharp cutoff for the 
regularization of the meson loops, the correction terms 
$\delta \Pi_\rho^{(k)\;\mu\nu, ab}$ are in general not transverse. 
However, as mentioned in \Sec{regularization}, we work in the
rest frame of the $\rho$-meson, i.e. ${\vec q}$~=~0.
In this particular case Eq.~(\ref{trans}) is not affected by the cutoff 
and the entire function ${\tilde \Pi}_\rho$ can be written in the form of 
Eq.~(\ref{pirho}):

\beq
    {\tilde \Pi}_\rho^{\mu\nu, ab}(q) \= 
    {\tilde \Pi}_\rho(q)\,T^{\mu\nu}\, \delta_{ab} \=
    \Big(\Pi_\rho(q) \+ \sum_{k}\; \delta \Pi_\rho^{(k)}(q) \Big)
    \,T^{\mu\nu}\, \delta_{ab}
    \;,
\label{pirho1}
\eeq
i.e. instead of evaluating all tensor components separately
we only need to calculate the scalar functions
$\Pi_\rho = -1/3\,g_{\mu\nu}\,\Pi_\rho^{\mu\nu}$ and 
$\delta\Pi_\rho^{(k)} = -1/3\,g_{\mu\nu}\,\delta\Pi_\rho^{(k)\,\mu\nu}$.  

A second consequence of vector current conservation is, that the 
polarization function should vanish for $q^2=0$.
For the correction terms this is violated by the sharp cutoff.
We cure this problem by performing a subtraction:
\beq
    \sum_{k}\; \delta \Pi_\rho^{(k)}(q) \;\rrr\;
    \sum_{k}\; \Big(\delta \Pi_\rho^{(k)}(q) \;-\;
    \delta \Pi_\rho^{(k)}(0) \Big) \;.
\label{sub}
\eeq 
Note, however, that already at the RPA level a subtraction is required,
although the RPA part is regularized by Pauli-Villars.
This is due to a rather general problem which is discussed in detail in
App.~\ref{correlators}.

\begin{figure}[t]
\hspace{1cm}
\parbox{14cm}{
     \epsfig{file=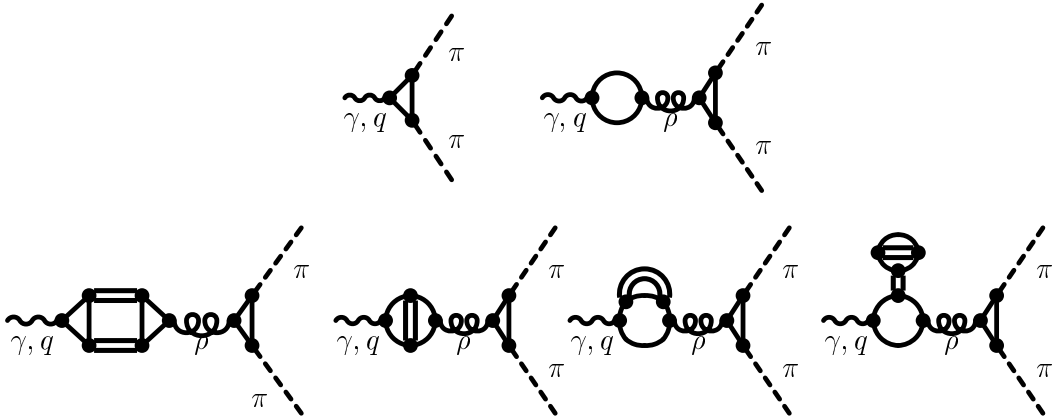,
     height=5.6cm, width=14.cm}}
\caption{\it Contributions to the pion electromagnetic form factor
         in the {\nce}. The propagator denoted by the curly line corresponds
         to the $1/N_c$-corrected $rho$-meson, while the double lines
         indicate RPA pions and sigmas.}
\label{figfeynpefm} 
\end{figure}

In Ref.~\cite{OBW} we have fixed $g_v$ and $\Lambda_M$ in the {\nce}, by fitting 
the pion electromagnetic form factor, $F_\pi(q)$, in the time-like region, 
which is dominated by the $\rho$-meson. The diagrams we included in that 
calculations are shown in Fig.~\ref{figfeynpefm}. 
The two diagrams in the upper part correspond to the standard
NJL description of the form factor~\cite{lutz} if the full
$\rho$-meson propagator (curly line) is replaced by the RPA one.
Hence, the first improvement is the use of the $1/N_c$-corrected 
$\rho$-meson propagator in the \nce. Since, in the
standard scheme, the photon couples to the $\rho$-meson via a quark-antiquark
polarization  loop, in the {\nce} we should also take into account the 
$1/N_c$-corrections
to the polarization diagram for consistency. This leads to the diagrams
in the lower part of Fig.~\ref{figfeynpefm}. On the other hand the external
pions are taken to be RPA pions (i.e. mass $m_\pi^{(0)}$ and 
pion-quark-quark coupling constant $g_{\pi qq}^{(0)}$). 
This is more consistent with the fact that the $\rho$-meson is also dressed 
by RPA pions and, as discussed above, we have fitted $m_\pi^{(0)}$ 
to the experimental value.

\begin{figure}[t]
\parbox{6cm}{
     \epsfig{file=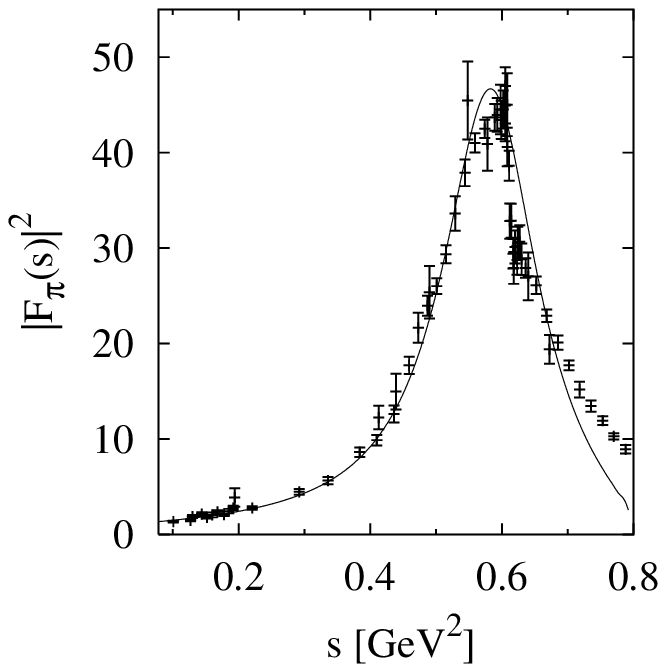,
     height=6cm, width=7.cm}\quad}
\hspace{2cm}
\parbox{6cm}{
     \epsfig{file=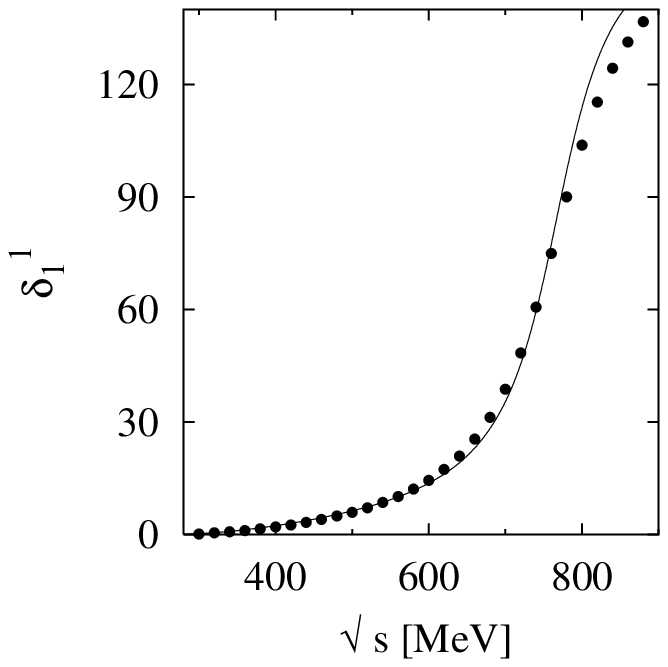,
     height=6cm, width=7cm}\quad}
\hspace{1cm}
\caption{\it The pion electromagnetic form factor (left panel) and 
             the $\pi\pi$-phase 
             shifts in the vector-isovector channel (right panel) for
             $\Lambda_M=600$~MeV and $g_v=1.6 g_s$. The other parameter values
             are taken from Table~\ref{tablence}. 
             The data points are taken from refs.~\cite{barkov} and
             \cite{frogatt}, respectively.}
\label{figpefm} 
\end{figure}

The numerical results for $|F_\pi|^2$ as a function of the center-of-mass
energy squared are displayed in the left panel of Fig.~\ref{figpefm}, 
together with the experimental data \cite{barkov}. 
The theoretical curve was calculated with a meson cutoff of 
$\Lambda_M=600$~MeV, a vector coupling constant $g_v = 1.6 g_s$ and the 
other parameters, $\Lambda_q$, $g_s$ and $m_0$ as listed in 
Table~\ref{tablence}. 
This roughly corresponds to a best fit to the data \cite{OBW}. 
Since we assumed exact isospin symmetry we can, of course,  
not reproduce the detailed structure of the form factor around 
0.61~GeV$^2$, which is due to $\rho$-$\omega$-mixing.  
The high-energy part above the peak is somewhat underestimated,
mainly due to the sub-threshold attraction in the $\rho$-mesons channel below
the $q\bar q$-threshold at $s$~=~0.80~GeV$^2$.
Probably the fit can be somewhat improved if we take a slightly 
larger meson cutoff, but we are not interested in fine-tuning here.

A closely related quantity is the charge radius of the pion,
which is defined as
\beq
     \ave{r_\pi^2} \= 6\,\frac{d F_\pi}{d q^2}\Big|_{q^2 = 0} \;.
\eeq 
With the above parameter set we obtain a value of 
$\ave{r_{\pi}^2}^{1/2} = .61$~fm. 
It lies slightly below the experimental value, 
$\ave{r_\pi^2}^{1/2}$~=~(0.663~$\pm$~0.006)~fm \cite{amendolia}. 
\begin{figure}[h!]
\hspace{3.5cm}
     \epsfig{file=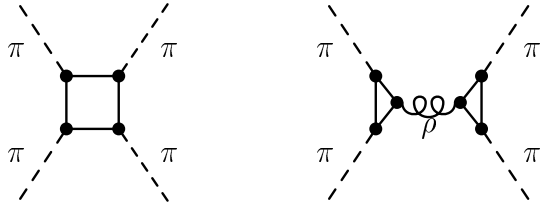}
\caption{\it Diagrams contributing to the $\pi\pi$-scattering 
             amplitude: Quark box diagram (left) and 
             s-channel $\rho$-meson exchange (right).}
\label{figfeynphase} 
\end{figure}

One can also look at the $\pi\pi$-phase shifts in the vector-isovector
channel. We include the diagrams shown in Fig.~\ref{figfeynphase}, i.e.
the s-channel $\rho$-meson exchange and the direct $\pi\pi$-scattering
via a quark box diagram. The latter has to be projected onto spin and
isospin~1, which is a standard procedure. (For example, the 
analogous projection onto spin and isospin~0 can be found in 
Refs.~\cite{davesne,bernard}.)  
The result, together with the empirical data \cite{frogatt}, is
displayed in the right panel of Fig.~\ref{figpefm}. 
Since the main contribution comes from the s-channel $\rho$-meson 
exchange, it more or less confirms our findings for the form factor:
below the $\rho$-meson peak a good of the data is obtained while, at higher
energies, where $q\bar q$-threshold 
effects start to play a role, we slightly overestimate the data.


\begin{figure}[h!]
\hspace{3.5cm}
     \epsfig{file=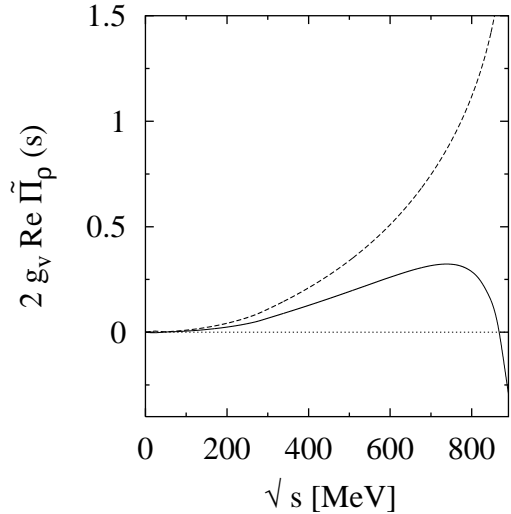}
\caption{\it Real part of the $\rho$-meson polarization function 
             $\tilde\Pi_\rho$ multiplied by $2 g_v$~= 17.6~GeV$^{-2}$
             as a function of the energy $\sqrt{s}$ in the rest frame
             of the meson. The dashed line corresponds to the {\nce}
             with $\Lambda_M$~=~600~MeV, the solid line to the LSS
             with $\Lambda_M$~=~700~MeV. The other parameters are given
             in Table~\ref{tablence} and Table~\ref{tablelss}, respectively.}
\label{figrhoselfsum} 
\end{figure}


Let us now turn to the LSS. As already discussed in the last paragraph 
of \Sec{secpi2}, there is not much room to vary the meson cutoff 
$\Lambda_M$ in this scheme: On the one hand $\Lambda_M$ is restricted
to values $\lsim$~700~MeV by the fit to $f_\pi$ and $\qq$ (see
Table~\ref{tablelss}). On the other hand we only have a chance to
get a realistic description of the $\rho$-meson if the constituent
quark mass $m$ is larger than at least 400~MeV. To achieve this, the
meson cutoff cannot be much smaller than 700~MeV. 
This means, $\Lambda_M$ is more or less fixed to this value and the
only remaining parameter is the vector coupling constant $g_v$.

It turns out, however, that with $\Lambda_M$~700~MeV we run already 
into instabilities in the $\rho$-meson channel. These instabilities
are of the same type as the instabilities in the pion channel,
(see \Sec{secpi1}), but unfortunately emerge already at lower values 
of $\Lambda_M$.   
This can be seen in \fig{figrhoselfsum} where the real part of the 
of the $\rho$-meson polarization function $\tilde\Pi_\rho$ multiplied
by $2g_v$ is plotted as a function of the energy $\sqrt{s}$ in the rest 
frame of the meson. The LSS result corresponds to the solid line.
For comparison we also show this function in the {\nce}, using the `best-fit 
parameters' given above (dashed curve).
  
According to \eqs{dm1} and {\ref{dmtilde}} the function 
$2g_v\,Re\tilde\Pi_\rho$ has to become equal to 1 for $\sqrt{s} \simeq m_\rho$,
crossing the line $2g_v\,Re\tilde\Pi_\rho = 1$ from below.
This is obviously the case in the {\nce}. In this scheme, 
$Re\tilde\Pi_\rho$ is a rising function and the above condition can be
easily fulfilled with the appropriate choice of $g_v$.
The situation is quite different in the LSS. Here $Re\tilde\Pi_\rho$
has a maximum at $\sqrt{s} \sim$~740~MeV and then steeply drops.
Hence, if $g_v$ is too small, the equation $2g_v\,Re\tilde\Pi_\rho$~=~1
has no solution at all (see \fig{figrhoselfsum}). On the other hand,
for large values of $g_v$ we get a ``phsyical'' solution at lower 
energies and an ``unphysical'' solution at higher energies. 
It is clear that none of these two scenarios would lead to a realistic
description of the $\rho$-meson. 


\begin{figure}[h!]
\parbox{6cm}{\epsfig{file=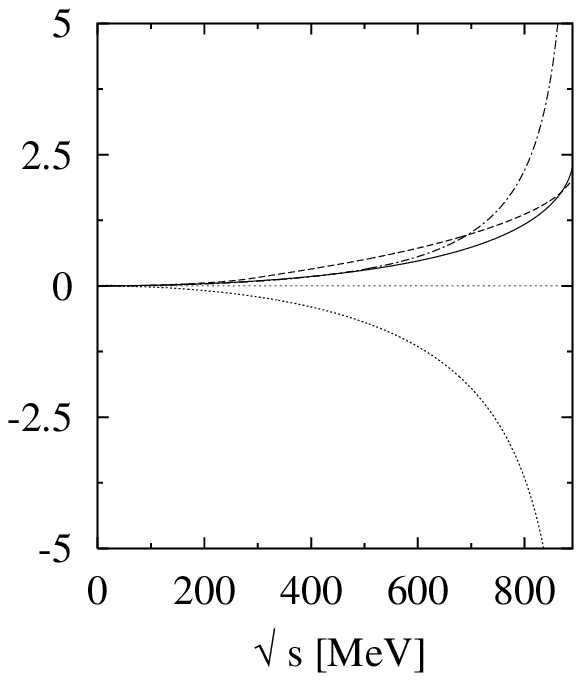}\quad}\hspace{2cm}
\parbox{6cm}{\epsfig{file=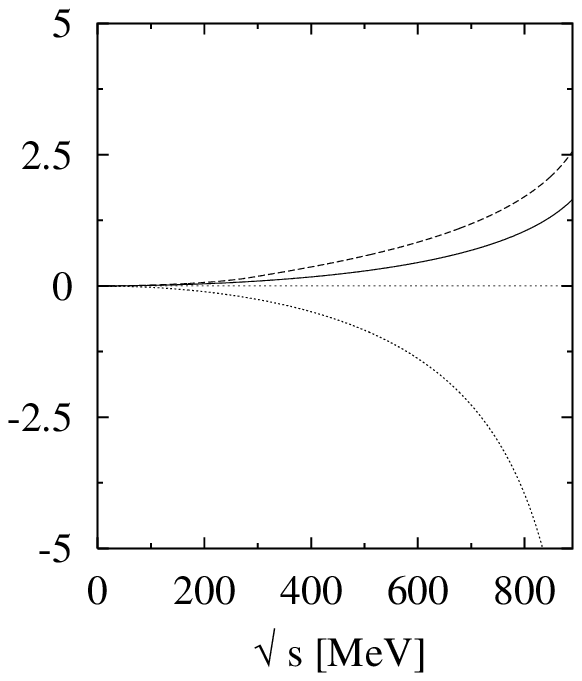}\quad}
\hspace{1cm}
\caption{\it RPA contribution $2g_v \Pi_\rho$ (solid) and the various
             correction terms to $2g_v \tilde\Pi_\rho$:
             $2g_v \delta\Pi_\rho^{(a)}$ (dashed),
             $2g_v (\delta\Pi_\rho^{(b)}+\delta\Pi_\rho^{(c)})$ (dotted)
             and $2g_v \delta\Pi_\rho^{(d)}$ (dashed-dotted).
             For all contributions we performed a subtraction, such
             that they vanish at $\sqrt{s}$~=~0.
             The left panel corresponds to the {\nce}, the right panel
             to the LSS. The model parameters are the same as in 
             \fig{figrhoselfsum}.} 
\label{figrhoself} 
\end{figure}


One might wonder, why the results in the {\nce} and in the LSS are so
different. To answer this question we have separately plotted the various 
contributions to the polarization function in \fig{figrhoself}. 
The left panel corresponds to the results in the {\nce}, the right panel
to the LSS. One immediately sees that the unphysical behavior in the
LSS is due the sum of the diagrams $\delta\Pi_\rho^{(b)}$ and 
$\delta\Pi_\rho^{(c)}$ (dotted), which is the only negative contribution. 
In the {\nce} these diagrams behave very similar. However, in this scheme 
their contribution is almost cancelled by the contribution of diagram
$\delta\Pi_\rho^{(d)}$, which is not present in the LSS. 

We should note that diagram $\delta\Pi_\rho^{(a)}$, which describes the
two-meson intermediate state is well-behaved in both schemes.
On the other hand the momentum dependence of all other diagrams is a
pure quark effect, which could be related to the imaginary part of
these diagrams above the (unphysical) two-quark threshold via
dispersion relations. Hence, if we could manage to further push up
the constituent quark mass, the momentum dependence of these contributions
should become smaller and the instabilities should eventually vanish.
Perhaps this is possible if further intermediate mesons, like 
$\rho$ and $a_1$ are included in the model.

\section{Quark condensate at $T\neq 0$}
\label{qqatt}
\setcounter{equation}{0}

It is expected, that at sufficiently large temperatures chiral symmetry, 
which is
spontaneously broken in vacuum, gets restored. The quark condensate as an
order parameter of chiral symmetry is well suited to study indications for
(partial) chiral symmetry restoration. At low temperatures model independent
results for the changes of the quark condensate can be obtained from
considering a gas of pions, which are the lightest particles and therefore
the main degrees of freedom
in this range. Approaching the phase transition we have to rely on model
calculations or lattice data because we do not have any fundamental knowledge
of the quark condensate at higher temperatures. Most of the results show a
phase transition at a temperature of $T_c \sim 150$ MeV. 

Among others, the NJL model has been used to examine the behavior of the quark
condensate as a function of temperature. Most of these investigations have been
performed in the mean-field approximation~\cite{klevansky,hatsuda,sklimt,mlutz}. There one finds a
second order phase transition with $T_c \sim 150-200$ MeV. However one has to
mention that these calculations suffer from the severe problem that the
thermodynamics is generated exclusively by a gas of quarks. One consequence
is that the low temperature behavior, which is driven mainly by pions,
is completely missed. Although we cannot by-pass the fundamental problem of
lack of confinement in the NJL model which in any case leads to the existence
of a quark gas at non-zero temperature, we can hope to improve the situation at
least at low temperatures via inclusion of mesonic degrees of freedom in a
calculation beyond mean-field. 

Therefore we begin with a closer look at the low-temperature behavior
of the quark condensate at $T\neq 0$.
After that we will discuss our numerical results within the 
$1/N_c$-expansion scheme and within the LSS.

\subsection{Low-temperature behavior}
\label{lowtemp}

In the chiral limit and at vanishing baryon density a
strict low-temperature expansion in chiral perturbation theory leads to the
following expression for the quark condensate \cite{Gasser}:
\beq
\qq_T = \qq \Big(1 - \frac{T^2}{8 f_{\pi}^2}-\frac{T^4}{384 f_{\pi}^4} +
\dots\Big)~.
\label{ChiPT}
\eeq 
Here $\qq$ denotes the quark condensate at zero temperature.
The $T^2$-term represents the contributions from a pure
pion gas, whereas the higher-order terms are due to interactions between the
pions. It has been shown \cite{Gasser} that the $T^2$- and the $T^4$-term
of this expansion are model independent results which follow from chiral
symmetry alone. Thus in principle every chirally symmetric model, including the
NJL model, should reproduce these terms. However, as $f_{\pi}$ is of the 
order $\sqrt{N_c}$, we can see that they are of the order $1/N_c$ 
and $1/N_c^2$, respectively. So a mean-field calculation, which corresponds
to a restriction to leading in $1/N_c$, will not be able to reproduce these 
terms \cite{florkowski}. Indeed, NJL model calculations in mean-field show 
a much more flat behavior at low temperatures
\cite{mlutz,ripka}:
\beq
\qq^{(0)}_T = \qq^{(0)} \Big(1- \frac{(2 m T)^{3/2}}{\pi^{3/2} \qq^{(0)}}
e^{-\frac{m}{T}} + \dots\Big)~.
\label{mf}
\eeq
Extending the calculations to next-to-leading order in $1/N_c$ will allow us
to reproduce the $T^2$-term. This will be demonstrated in the following. 

Our calculations at non-zero temperature are performed within imaginary
time formalism. Basically this amounts to replacing the energy integration 
in the various $n$-point functions by a sum over Matsubara frequencies.
The explicit expressions are listed in App.~\ref{aptemp}.
As there exists a preferred frame of reference in the heat bath, all dynamical
quantities depend separately on energy and three-momentum.
Hence in the following, a finite-temperature RPA propagator, for instance, 
will be denoted as $D_M(\omega,\vec p)$. For scalar quantities, like
masses or condensates at non-zero temperature we use a suffix $T$
in order to distinguish them from the analogous quantities in vacuum
(cf. \eqs{ChiPT} and (\ref{mf})).  

In analogy to the vacuum expressions (\eqs{qbq0} and (\ref{deltaqbqexp}))
the quark condensate in next-to-leading order of the $1/N_c$-expansion
scheme is given by
\beq
\qq_T = \qq^{(0)}_T + \delta\qq_T = -\frac{m_T-m_0}{2 g_s}-\frac{D_{\sigma}(0,0)
  \Delta_T}{2 g_s}~.
\label{qq}
\eeq
As shown in \eq{mf}, the leading-order term $\qq^{(0)}_T$ does not
contribute to the change of the quark condensate to order $T^2$.
Similarly, thermal effects in the $\sigma$-meson propagator can be neglected 
at low temperatures. 
Therefore we only need to consider the temperature dependence of $\Delta_T$. 
If standard techniques are used the sum over the Matsubara frequencies in
Eq.~(\ref{delta}) can be converted into a contour integral \cite{fetter}:
\bea
 \Delta_T 
&=& 4i N_c N_f\ m_T \frac{1}{2\pi i}
\int\frac{d^3p}{(2\pi)^3}\int_C\frac{dz}{e^{z/T}-1}{\Big\{}
D_{\pi}(z,\vec{p})\  (3I(0,0)-3 (z^2-\vec{p}^2)\
K(z,\vec{p}))\nonumber \\
&&\hspace{2.5cm} +  D_\sigma(z,\vec{p})\ (2\ 
I(z,\vec{p})+I(0,0)-(z^2-\vec{p^2}-4 m_T^2) \ K(z,\vec{p})) {\Big\}}~. 
\label{contour}
\eea
At low temperatures,
the main contribution to the temperature-dependent part of this integral comes
from the lowest lying pion pole, as the other contributions are exponentially
suppressed. In the chiral limit we can therefore approximate this part for low
temperatures by
\beq
 \Delta_T - \Delta
= 4 N_c N_f\ m 
\int\frac{d^3p}{(2\pi)^3}\frac{2}{e^{|\vec{p}|/T}-1}{\Big\{}
\frac{3}{2|\vec{p}| 2 N_c N_f} {\Big\}}~.
\eeq
This integral can be evaluated analytically and we obtain:
\beq
 \Delta_T-\Delta \= m\,{\frac{T}{2}}^2~.
\eeq
The last step is to realize that in the chiral limit the vacuum
$\sigma$-meson propagator can be expressed through the leading-order
pion decay constant as
\beq
    D_{\sigma}(0) \= -\frac{1}{4f_{\pi}^{2(0)}}
\eeq
(see \eqs{sigmai} and (\ref{fpiex})).
We finally obtain for the quark condensate in next-to-leading order at low
temperatures:
\beq
\qq_T = \qq- \qq^{(0)}\frac{T^2}{8f_{\pi}^{2(0)}}  
\label{qqt}
\eeq
Comparing this with the chiral perturbation theory result, \eq{ChiPT},
we see that we can in principle reproduce the $T^2$-term. 
Note, however, that the coefficient is given by the quark condensate
and the pion decay constant in leading order in $1/N_c$, 
according to a strict expansion of \eq{ChiPT} up to next-to-leading order
in $1/N_c$. 
The physical reason for this behavior is the fact that the $1/N_c$
corrections to the quark condensate correspond to fluctuating RPA-mesons 
and hence the thermal corrections at low temperatures are due to thermally 
excited RPA pions in this model. 

For the LSS, a similar result has been derived in Ref.~\cite{florkowski}.
In the chiral limit the authors find 
\beq
\qq_T = \qq \Big(\,1 \;-\; \frac{T^2}{8f_{\pi}^{2(0)}}\,\Big)~.  
\label{qqtlss}
\eeq
Here $f_{\pi}^{2(0)}$ is understood as the RPA-pion decay constant,
\eq{fpiex}, but evaluated at the quark mass $m$, which follows from
the LSS gap equation, \eq{localgap}. 
This corresponds to the fact that in the LSS the thermal corrections
to the quark condensate at low temperatures are due to RPA pions which 
consist of LSS quarks.

 
\subsection{Numerical results within the $1/N_c$-expansion scheme}
\label{tnc}

\begin{figure}[b!]
\parbox{6cm}{
     \epsfig{file=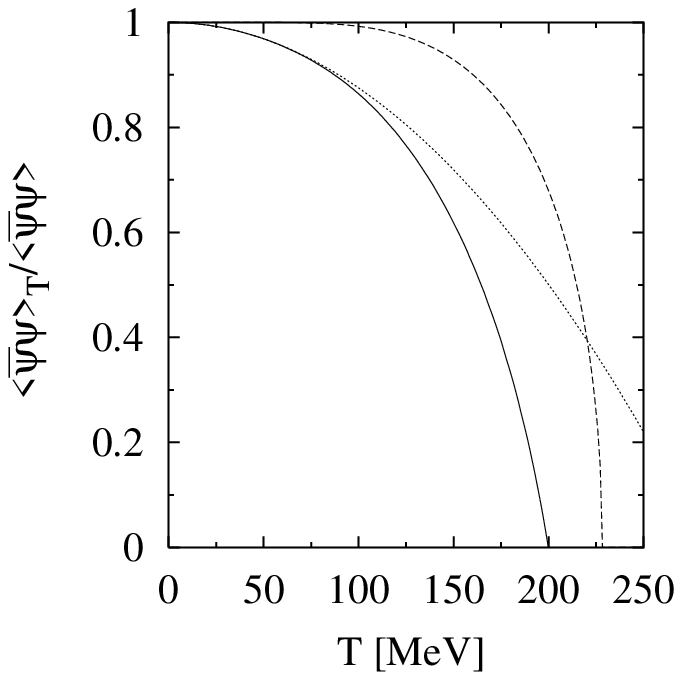,
     height=6cm, width=7.cm}\quad}
\hspace{2cm}
\parbox{6cm}{
     \epsfig{file=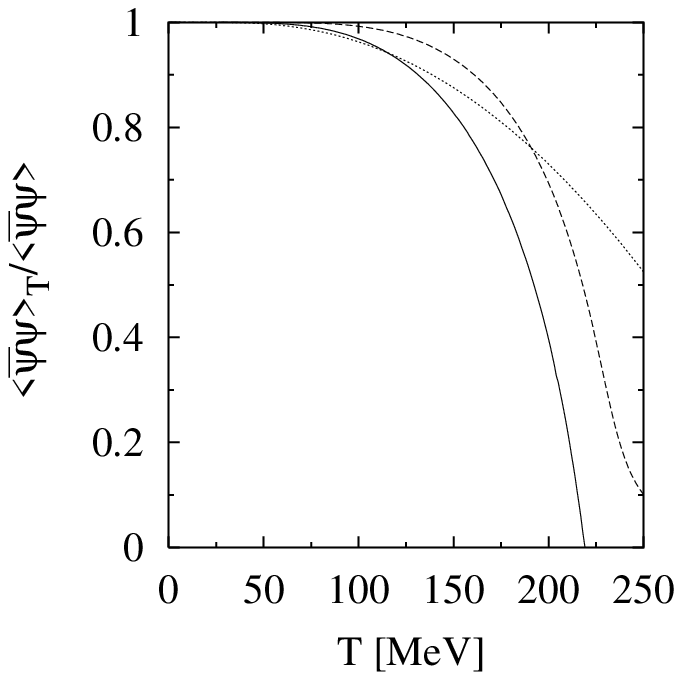,
     height=6cm, width=7cm}\quad}
\hspace{1cm}
\caption{\it Quark condensate as a function of temperature, normalized to the
     vacuum value, in the chiral limit (left) and with $m_{\pi}^{(0)} = 140$
     MeV (right). Leading order in
     $1/N_c$ (dashed line), next-to-leading order (solid line) and free
     pion gas (dotted line).}
\label{fig10} 
\end{figure}

Our numerical results for the temperature behavior of the quark condensate 
within the $1/N_c$-expansion scheme are displayed in Fig.~\ref{fig10}.
The r.h.s. corresponds to a realistic parameter set with 
$m_\pi^{(0)}$~=~140~MeV (Table~\ref{tablence} with $\Lambda_M =$~600~MeV), 
the l.h.s. to the chiral limit. 
The solid lines indicate the results obtained in next-to-leading order.
For comparison we also show the leading order (dashed line) and the pure 
pion gas result (dotted).  

We begin our discussion with the chiral limit.
At low temperatures ($T \lsim 100$~MeV) our results show the behavior 
discussed in the previous subsection: The next-to-leading order result 
is in very good agreement with the pion gas result (Eq.~\ref{qqt}), 
whereas the leading-order result remains almost constant.
Therfore in this regime the extention of the NJL model to next-to-leading order 
in $1/N_c$ leads to a considerable improvement.
Since the unphysical quark degrees of freedom, which are in principle
always present in the NJL model, are exponentially suppressed, 
the system is dominated by the (physical) pion degrees of freedom, 
which come about in next-to-leading oder. 

However, because of the much larger degeneracy factor (24 as compared to 3)
we cannot avoid that effects due to thermally excited quarks become important
at some temperature. In our present calculation this happens at about 
$T\sim 100$~MeV. In a free gas approximation, this roughly corresponds to 
the temperature, at which the quark pressure becomes equal to the pion 
pressure.

At this point one might raise the question about the physical meaning
of quark effects at these temperatures.
In nature, quark degrees of freedom can only be excited above the
deconfinement phase transition. 
In the NJL model there is no confinement and hence no deconfinement
transition. However, lattice calculations \cite{laermann} 
indicate that the deconfinement phase transition at finite temperature
coincides with the chiral phase transition.
One should therefore compare the temperature at which thermally
excited quarks become important with the critical temperature for
the chiral phase transition. 
From the above point of view, quark effects below the phase transition 
are either invisible or unphysical.
On the other hand, at least close to the phase transition one might relax 
this strict position.  
In this regime one might think of a resonance gas with many degrees of 
freedom, which could be effectively described by a quark gas 
(``quark-hadron duality'').

Unfortunately, as already pointed out in \Sec{nonpert}, the perturbative 
treatment of the mesonic fluctuations does not allow for a description of 
the chiral phase transition. Although the quark condensate vanishes at
$T\sim 200$~MeV, this does not correspond to a true phase transition.
(Note that the slope of the curve does not diverge at this point.)
In any case, the applicability of the perturbative expansion scheme 
probably breaks down much earlier. Therefore we cannot give a definite
answer to the question whether the thermally excited quarks become
important near the phase transition or much below. 

Our results  with $m_0 \neq 0$ are shown on the r.h.s. of Fig.~\ref{fig10}.
Since the RPA pions are now massive and therefore exponentially suppressed,
the quark condensate as a function of $T$ stays much more flat than in
the chiral limit. Nevertheless, at low temperatures pions can still be
most easily excited as they are the lightest particles.
Therefore the next-to-leading order result (solid line) can be approximated
quite well albeit not perfectly by the pure pion gas result (dotted) 
in this regime.
The latter was calculated from the pressure $p_\pi$ of a massive
pion gas as
\beq
\qq_T \= \qq \+ \qq^{(0)}\,\frac{m_0}{f_{\pi}^{2(0)}}\,
         \frac{dp_\pi(T)}{dm_\pi^{2\,(0)}}  
\label{qqtm}
\eeq
which can be easily derived with the help of the GOR relation.

Quark effects become important at almost the same temperature as in the 
chiral limit, at $T\sim100$ MeV.

 
\subsection{Local selfconsistent scheme}
\label{tlss}

\begin{figure}[t!]
\parbox{6cm}{
     \epsfig{file=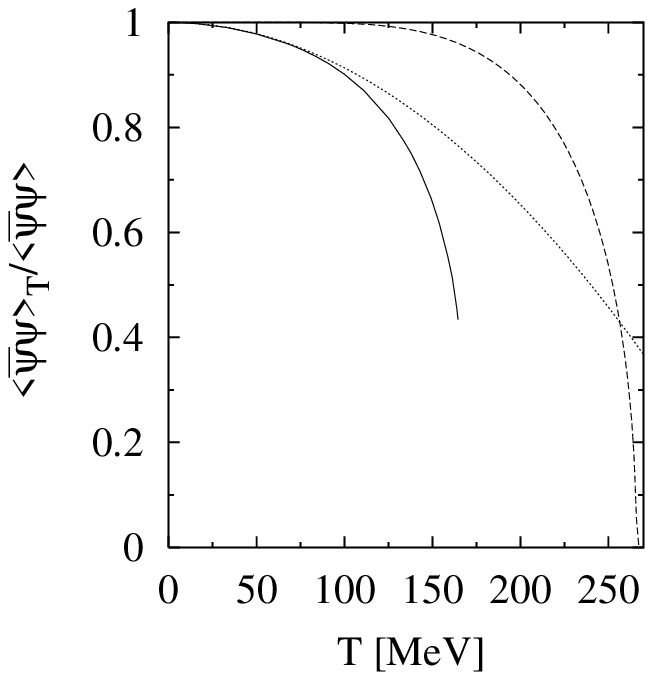,
     height=6cm, width=7.cm}\quad}
\hspace{2cm}
\parbox{6cm}{
     \epsfig{file=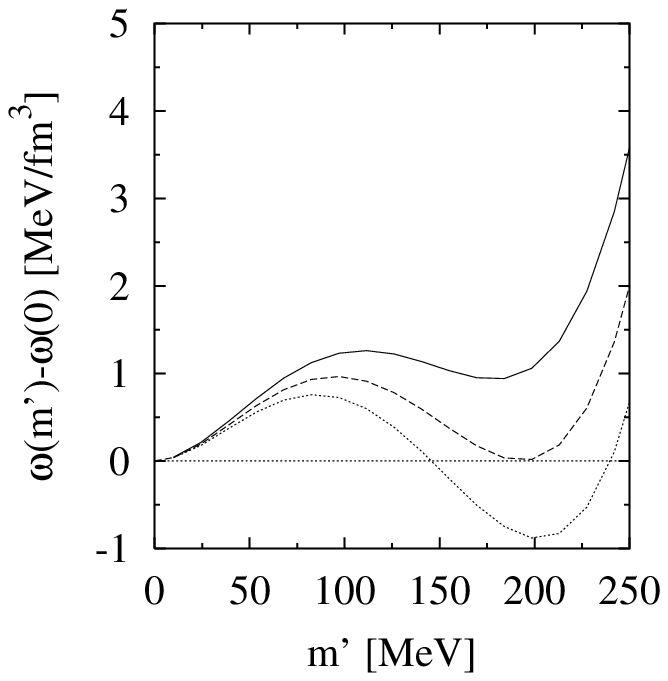,
     height=6cm, width=7cm}\quad}
\hspace{1cm}
\caption{\it Left: Quark condensate in the chiral limit as a function of
     temperature, normalized to the vacuum value, Hartree approximation 
     (dashed), local selfconsistent scheme (solid) and free
     pion gas (dotted). 
     Right:
     Thermodynamic potential per volume as a function of the constituent 
     quark mass in the selfconsistent scheme with $T=163.9$ MeV (dotted), 
     $T=164.5$ MeV (dashed) and $165.3$ MeV (solid).}
\label{fig11} 
\end{figure}

Let us now compare the results of the previous subsection with the 
analogous calculations in the LSS. 
A study of the temperature dependence of the quark 
condensate within the LSS can also be found in Ref.~\cite{florkowski}. 
Here we restrict ourselves to the chiral limit. 

Our results are shown on the l.h.s. of \fig{fig11}. 
The calculations have been performed using the parameters of 
Table~\ref{tablelss} for $\Lambda_M = 700$~MeV, but $m_0 = 0$.
As discussed in \Sec{lowtemp}, at low temperatures the model behaves
again like a free pion gas (dotted line). 
Deviations from this behavior become visible at $T\sim 100$~MeV which 
is quite similar to our observations in the $1/N_c$-expansion scheme.

In contrast to the $1/N_c$-expansion scheme, the treatment of the
mesonic fluctuations in the LSS also allows an examination of the
phase transition. With the present parameters it takes place at
$T_c = 164.5$~MeV which is considerably lower than in Hartree approximation,
where we have $T_c= 266.1$~MeV. 
Note, however, that about one third of this 
reduction can be attributed to the fact that the constituent quark mass 
$m$~=~468.4~MeV in the LSS is lower than the corresponding Hartree mass 
$m_H$~=~600.0~MeV. For $m_H$~=~468~MeV we would get a critical temperature 
of about 236~MeV in Hartree approximation. 
(It is also interesting to note, that the critical temperature in the 
LSS calculation almost coincides with the critical temperature 
$T_c^{RPA}$~=~164.4~MeV one obtains in Hartree approximation for the 
parameters fitted in the RPA, i.e. $m = m_H$~=~260~MeV.) 

Whereas in Hartree approximation the phase transition is of 
second order, in the LSS the system undergoes a first-order phase transition,
as already reported in Ref.~\cite{florkowski}.
This can be inferred from the thermodynamic potential $\omega$,
which is displayed on the r.h.s. of \fig{fig11} for different temperatures
as a function of the constituent quark mass $m'$. At $T = 164.5$~MeV one
can clearly identify two degenerate minima at $m' = 0$ and $m' \neq 0$,
corresponding to a first-order phase transition at that temperature.
One can ask whether this phenomenon depends on the strength of
the mesonic fluctuations which can be controlled by the cutoff
$\Lambda_M$. Varying this parameter we find that the discontinuity
decreases with decreasing $\Lambda_M$, but even for very small values 
of the cutoff we encounter a first order phase transition.

Let us come back to the questions about the relevance of the unphysical
quark degrees of freedom. As already mentioned, deviations from the pure 
pion gas result become visible at $T\sim 100$~MeV, which corresponds to 
about $0.6 T_c$. At this temperature one would not expect quark effects
to be present in nature.
Furthermore, according to universality arguments, it is generally believed,
that the finite-temperature chiral phase transition in QCD with two massless
quarks is of
second order \cite{pisarski}. This is based on the assumption that
at $T_c$ there are four massless bosonic degrees of freedom (three pions
and one $\sigma$) which determine the infrared behavior of the system.
In this case QCD -- but also the NJL model -- should lie in the same 
universality class as the $O(4)$-model, which is known to have a 
second-order phase transition. 
Although some time ago it was claimed, that this argument might not hold
if the boson fields are not elementary but composite \cite{kocic},
it is probably more likely that the first-order phase transition we
observe is an artifact of the approximation scheme.
In this context the application of renormalization group techniques to 
the NJL model would be extremely interesting.

\section{Conclusions}
\label{conclusions}
\setcounter{equation}{0}

We have investigated quark and meson properties within the 
Nambu--Jona-Lasinio model, including meson-loop corrections. These have 
been generated in two different ways. The first method is a systematic 
expansion of the self-energies in powers of $1/N_c$ up to next-to-leading 
order \cite{dmitrasinovic, oertel, OBW}. 
In the second scheme, a local correction term to the standard Hartree 
self-energy is self-consistently included in the gap equation 
\cite{dmitrasinovic}. 
We therefore call it the ``local selfconsistent scheme'' (LSS).
This scheme can also be derived as the one-meson-loop approximation 
to the effective action \cite{nikolov}. 
Both schemes, the {\nce} and the LSS, are consistent with chiral symmetry, 
leading to massless pions in the chiral limit. For non-vanishing current 
quark masses the pion mass is consistent with the Gell-Mann--Oakes--Renner 
relation in the LSS. This is also true in the {\nce} if one carefully expands 
both sides of the relation up to next-to-leading order in $1/N_c$.

The relative importance of the mesonic fluctuations is controlled by a 
parameter $\Lambda_M$, which cuts off the three-momenta of the meson loops. 
In both schemes we encounter instabilities in the pion propagator if the 
meson effects become too strong. 
In order to find out whether these instabilities are related to an
unstable ground state \cite{oertel,kleinert},
leading to a ``chiral restoration phase transition''
at some critical value of $\Lambda_M$, we calculated the effective
action of the LSS for increasing values of $\Lambda_M$.  
(Note, that such investigations are not possible within the 
{\nce}, where mesonic fluctuations are included only perturbatively.)
It turned out, that up to a certain value of $\Lambda_M$ the system indeed 
seems to move towards a ``phase transition''. However, 
when $\Lambda_M$ is further increased the non-trivial ground state becomes 
again more stable and no phase transition takes place.

Of course, at the end, the value of $\Lambda_M$, together with the other
parameters, has to be determined by fitting physical observables.  
The $\rho$-meson and related quantities are very well suited for this purpose,
since the meson loops are absolutely crucial in order to include the 
dominant $\rho \rightarrow \pi\pi$-decay channel, while the Hartree+RPA 
approximation contains only unphysical $q\bar q$-decay channels.
Here another problem, which constraints the possible choice of parameter
values, becomes obvious: A priori it is not clear to what extent these
unphysical decay modes, which are an unavoidable consequence of the missing
confinement mechanism in the NJL model, are still present in the region 
of the $\rho$-meson peak. 

For the {\nce}, the parameters have already been fixed in Ref.~\cite{OBW}.
We obtained a reasonable fit of $f_{\pi}$, $\qq$ and the pion electromagnetic 
form factor with a constituent quark mass of $m$~=~446~MeV. This means,
the unphysical $q\bar q$-decay channel opens at 892~MeV, about 120~MeV above 
the maximum of the $\rho$-meson peak. Furthermore, the parameters of that 
fit are far away from the region, where the instabilities in the pion 
propagator emerge. In fact, we found only moderate changes in the pion and
quark sector: $f_\pi$ and $\qq$ are lowered by about 20\% by the meson loop
corrections, while the pion mass is increased by about 10\%. This indicates 
that the $1/N_c$ expansion converges rapidly and higher-order
terms in the $1/N_c$-expansion are small. 

Unfortunately we did not succeed to obtain a similar fit within the LSS.
Since in this scheme the meson-loop effects lower the constituent quark mass  
as compared to the Hartree mass, it is much more difficult to evade the 
problem of unphysical $q\bar{q}$-decay channels in the vicinity of the 
$\rho$-meson peak. We found that a relatively large meson cutoff,
$\Lambda_M \sim$~700~MeV is needed in order to get the quark mass large
enough and at the same time a fit for $f_\pi$.
However, to our surprise for this cutoff the $\rho$-meson self-energy
already suffers from stability problems, similar to those already discussed
for the pion. As a result we are not able to get a reasonable description
of the $\rho$-meson propagator and hence of the pion electromagnetic 
form factor within the LSS.
It remains to be checked, whether these problems can be cured by taking 
into account additional intermediate states, like vector mesons and axial 
vector mesons or by different way of regularization.

In the last part of this article we have investigated the temperature dependence 
of the quark condensate. 
In both schemes the low-temperature behavior is consistent with lowest-order
chiral perturbation theory, i.e. the temperature dependence arising from a
free pion gas. This is a considerable improvement over the mean-field result,
where the temperature dependence is entirely due to thermally excited
quarks, i.e. unphysical degrees of freedom.  
At higher temperatures, however, thermal quark effects also become visible 
in the two extended schemes. We argued that this could be tolerable 
near the chiral phase boundary which is, according to lattice results,
identical to the deconfinement phase boundary at non-zero temperatures.

Whereas the perturbative treatment of the mesonic fluctuations within the 
{\nce} does not allow an examination of the chiral phase transition, this 
is possible in the LSS. 
For our model parameter set we found a critical temperature of 164.5~MeV.
On the other hand, quark effects are visible already at a temperature of 
$\sim$~100~MeV. Obviously this is still too early to be realistic.
Maybe here the model can be improved by including additional intermediate 
meson states.

In agreement with Ref.~\cite{florkowski}
we found a first-order phase transition in that scheme. This contradicts
the general belief that the non-zero temperature chiral phase transition 
in a model with two light flavors should be of second order and is 
probably an artifact of the approximation. Here further investigations,
e.g. applying renormalization group techniques, would be very interesting.


\section*{Acknowledgments}
We are indebted to G.J. van Oldenborgh for his assistance in questions
related to his program package FF (see  
{\it http://www.xs4all.nl/$\sim$gjvo/FF.html}),
which was used in parts of our numerical calculations.
We also thank G. Ripka, B.-J. Schaefer and M. Urban for illuminating 
discussions. 
This work was supported in part by the BMBF and NSF grant NSF-PHY98-00978.


\begin{appendix}


\section{Definition of elementary integrals}
\label{integrals}
\setcounter{equation}{0}

It is possible to reduce the expressions for the quark loops to some
elementary integrals~\cite{passarinoveltman}, see App.~\ref{correlators}
and~\ref{functions}. In this section we give the definitions of these 
integrals. 
\bea
&&\hspace{-12mm} I_1 = \intk \frac{1}{k^2-m^2+i \eps}~, \label{onepoint}\\
&&\hspace{-12mm} I(p) = \intk \frac{1}{(k^2-m^2+i \eps)( (k+p)^2-m^2+
  i\eps)}~,\label{twopoint}\\
&&\hspace{-12mm} K(p) = \intk \frac{1}{(k^2-m^2+i \eps)^2( (k+p)^2-m^2+
  i\eps)}~,\label{threepoint1} \\
&&\hspace{-12mm} M(p_1,p_2) = \intk \frac{1}{(k^2-m^2+i \eps)
                 ( k_1^2-m^2+ i\eps)( k_2^2-m^2+i \eps)}~,\label{threepoint}\\
&&\hspace{-12mm} L(p_1,p_2,p_3) = \intk\frac{1}{(k^2-m^2+i \eps)
  ( k_1^2-m^2+ i\eps)
  ( k_2^2-m^2+i \eps)( k_3^2-m^2 + i\eps)}~,\label{fourpoint}\\
&&\hspace{-12mm} p_1^{\mu} M_1(p_1,p_2)+p_2^{\mu} M_1(p_2,p_1)
  = \intk \frac{k^{\mu}}{(k^2-m^2+i \eps)( k_1^2-m^2+ i\eps)
  (k_2^2-m^2+i\eps)},\label{tensor}
\eea
with $ k_i = k+p_i$. 
The function $M_1(p_1,p_2)$ can be expressed in terms of the other integrals:
\beq
M_1(p_1,p_2)
= \frac{p_1\!\cdot\!p_2\ I(p_1) - p_2^2\ I(p_2)
   + (p_2^2-p_1\!\cdot\!p_2)\  I(p_1-p_2)  
  + p_2^2\ (p_1^2-p_1\!\cdot\!p_2)\ M(p_1,p_2)}
  {2 \; ((p_1\!\cdot\!p_2)^2 \,-\, p_1^2\,p_2^2)}~,
\eeq
All integrals in Eqs.~(\ref{onepoint}) to (\ref{tensor}), are understood 
to be regularized. As described in \Sec{regularization} we use 
Pauli-Villars regularization with two regulators, i.e. we replace
\beq
    \intk f(k;m) \;\rrr\; \intk \sum_{j=0}^2 c_j\,f(k;\mu_j)~,
\eeq
with
\beq
    \mu_j^2 \= m^2 \+ j\,\Lambda_q^2~;  \qquad
    c_0 = 1, \quad c_1 = -2, \quad c_2 = 1~.
\eeq
One then gets the following relatively simple analytic expressions for 
the integrals $I_1$, $I(p)$ and $K(p)$: 
\bea
&&\hspace{-12mm} 
  I_1=\frac{-i}{16\pi^2}\sum_j c_j\, \mu_j^2 \ln(\mu_j^2)\label{i1}\\
&&\hspace{-12mm}
  I(p) = \frac{-i}{16\pi^2}\sum_j c_j\, \Big(x_{j1} \ln(x_{j1})
  +x_{j2} \ln(-x_{j2})+x_{j1}\ln(-p^2x_{j1})+x_{j2}\ln(p^2x_{j2})\Big)\\ 
&&\hspace{-12mm}
  I(p=0) = \frac{-i}{16\pi^2}\sum_j c_j\, \ln(\mu_j^2)\label{i0}\\
&&\hspace{-12mm}
  K(p) = \frac{-i}{16\pi^2}\sum_j c_j\,\frac{1}{2 p^2(x_{j1}-x_{j2})} 
\Big(-\ln(x_{j1})-\ln(-x_{j1})+ \ln(x_{j2})+\ln(-x_{j2})\Big)~,
\eea
with 
\beq
x_{j1,2} = {1\over2}\pm{1\over2}\sqrt{1-{4 \mu_j^2\over p^2}} \;.
\eeq
An analytic expression for the three-point function (Eq.~\ref{threepoint})
can be found in Refs.~\cite{vanoldenborgh} and~\cite{veltman}. 
In certain kinematical regions the four-point function (eq.~\ref{fourpoint}) 
is also known analytically \cite{vanoldenborgh,veltman}.


\section{RPA propagators}
\label{correlators}
\setcounter{equation}{0}

Using the definitions given in the previous section the gap equation
(Eq.~(\ref{gap})) takes the form
\beq
    m \= m_0 + 2ig_s\ 4 N_c N_f\ m\ I_1 \;.
\label{gapex}
\eeq
Similarly one can evaluate the quark-antiquark polarization diagrams 
(Eq.~(\ref{pol0})) and calculate the RPA meson propagators.
The results for $\sigma$-meson and pion read: 
\bea
  D_\sigma(p) &=& \frac{-2 g_s}{1-2i g_s\ 2 N_c N_f\ 
 (2 I_1 - (p^2-4m^2)\ I(p))}~,\label{sigma}\\   
  D_{\pi}(p) &=& \frac{-2 g_s}{1- 2i g_s\ 2 N_c N_f\ (2 I_1-p^2\ I(p))}~.
 \label{pseudopi} \   
\eea
If we evaluate these propagators with the constituent
quark mass in Hartree approximation we can simplify the above expressions with
the help of the gap equation (Eq. \ref{gapex}) to obtain:
\bea   
  D_\sigma(p) &=& \frac{-2 g_s}{{m_0\over m} +2i g_s\ 2 N_c N_f\ 
 (p^2-4m^2)\ I(p)}~,\label{sigmai}\\   
  D_{\pi}(p) &=& \frac{-2 g_s}{\frac{m_0}{m}+2i g_s\ 2 N_c N_f\ p^2\ I(p)}~.
 \label{pseudopii}
\eea
As discussed in \Sec{solution}, this form is also used for the internal
meson propagators in the LSS.

A straight-forward evaluation of the vector and axial vector polarization 
diagrams gives
\bea
    \Pi_\rho(p) &=&  -i {4\over3} N_c N_f\ (-2 I_1+(p^2+2 m^2)\ I(p))~,
    \label{pirhoex}\\
    \Pi_{a_1}(p) &=&  -i {4\over3} N_c N_f\ (-2 I_1+(p^2-4 m^2)\ I(p))~.
    \label{pia1ex}
\eea
Because of vector current conservation $\Pi_\rho$ should vanish
for $p^2$~=~0. This is only true if 
\beq
     m^2\,I(0) \= I_1~,
\label{mi0}
\eeq
which is not the case if we regularize $I(p)$ and $I_1$ as described
in App.~\ref{integrals}. This corresponds to the standard form of
Pauli-Villars regularization in the NJL model \cite{klevansky}. 
Alternatively one could perform the replacement Eq.~(\ref{pv}) for 
the entire polarization loop. In fact, this is more in the original
sense of Pauli-Villars regularization \cite{itzykson}. Then the factor 
$m^2$ in Eq.~(\ref{pirhoex}) should be replaced by a factor $\mu_j^2$ 
inside the sum over regulators and one can easily show that Eq.~(\ref{mi0}) 
holds (see~Eqs.~(\ref{i1}) and~(\ref{i0})). 
However, this scheme would lead to even more severe problems:
From the gap equation (Eq.~\ref{gapex}) we conclude that $i I_1$ should 
be positive. On the other hand the pion decay constant in the chiral limit
and in leading order in $1/N_c$ is given by \cite{klevansky}
\beq
f_{\pi}^{2(0)} \=  -2i N_c N_f\ m^2\ I(0)~.
\label{fpiex}
\eeq
which implies that $i m^2 I(0)$ should be negative.
So irrespective of the regularization scheme Eq.~(\ref{mi0}) cannot be 
fulfilled if we want to get reasonable results for $m$ and $f_\pi^{(0)}$
at the same time. 
Therefore we choose the standard form of Pauli-Villars regularization in the 
NJL-model \cite{klevansky} and replace the term $I_1$ in 
Eq.~(\ref{pirhoex}) by hand by $m^2\ I(0)$. For consistency the $a_1$ is
treated in the analogous way. This leads to the following $\rho$- and
$a_1$-meson propagator 
\bea
  D_\rho(p) &=& \frac{-2 g_v}{1+ 2i g_v\ {4\over3} N_c N_f\ 
 (-2 m^2\ I(0)+(p^2+2 m^2)\ I(p))}\label{rhoi}~,\\   
  D_{a_1}(p) &=& \frac{-2  g_v}{1+ 2i g_v\ {4\over3} N_c N_f\ (-2 m^2\
    I(0)+(p^2-4 m^2)\ I(p))}~.\label{a1i}
\eea   


\section{Explicit expressions for the meson-meson vertices}
\label{functions}
\setcounter{equation}{0}

In this section we list the explicit formulae for the meson-meson 
vertices. We restrict ourselves to those combinations which are needed
for the calculations presented in this article.

We begin with the three-meson vertices $\Gamma_{M_1,M_2,M_3}(q,p)$
(see~Fig.~\ref{fig4}): 
\bea
-i\Gamma_{\sigma,\sigma,\sigma}(q,p) &=& i 2 m N \Big(I(p^{\prime})
          +I(q)+I(p)+(4m^2-\frac{1}{2}(p^{\prime2}+p^2+q^2)) 
          M (p,-q)\Big) ~,
\nonumber\\
-i\Gamma^{ab}_{\pi,\pi,\sigma}(q,p) &=& i 2 m N\delta_{ab} 
\Big(I(p^{\prime})+p\!\cdot\!q M(p,-q)\Big) ~,
\nonumber\\
-i\Gamma^{\mu\lambda,ab}_{\rho,\rho,\sigma}(q,p)
&=&\delta_{ab}h(q,p)\Big(g^{\mu\lambda} 
- \frac{p^2\,q^{\mu}q^{\lambda} \+ q^2\,p^{\mu}p^{\lambda}
        \;-\;p\!\cdot\!q\,(p^{\mu}q^{\lambda}+q^{\mu}p^{\lambda})}
        {p^2q^2 \;-\; (p\!\cdot\!q)^2} \Big)~,
\nonumber\\ 
h(q,p) &=& i m N\Big(I(q)+I(p)-2 I(p^{\prime})
+(4 m^2-2 p\!\cdot\!q-p^{\prime2}) M(p,-q)\Big)~,
\nonumber\\
-i\Gamma_{\pi,\pi,\rho}^{\mu,abc}(q,p) &=&  \eps_{abc} \Big(q^{\mu}
f(q,p)-p^{\mu} f(p,q)\Big)~, \nonumber\\
f(q,p) &=& N \Big(-I(q)+p^2 M(p,-q)+2 p\!\cdot\!q M_1(q,-p)\Big)~,
\eea
with $p^{\prime}= -p-q$ and $N=4 N_c N_f$.

For the four-meson vertices we only need to consider the special cases 
needed for the diagrams (b) and (c) in Fig.~\ref{fig3}:
\bea
-i\Gamma_{\sigma,\sigma,\sigma,\sigma}(q,p,-q)&=& -N \Big\{
\frac{I(p-q)+I(p+q)}{2} + 4 m^2\ (M(p,q)+M(p,-q))
\nonumber\\ && \hspace{1cm} 
+ 2 \big(m^2\ (4 m^2-p^2-q^2)-\frac{p^2q^2}{4}\big)\
L(p,-q,p-q)\Big\}
\nonumber\\
-i\Gamma_{\sigma,\sigma,\sigma,\sigma}(q,p,-p) &=& -N \Big\{ I(p+q) + I(0)
+ 4m^2 \big( K(p)+K(q)+2 M(p,-q)\big) 
\nonumber\\ && \hspace{1cm} 
+2 p\!\cdot\!q M(p,-q) -q^2 K(q)-p^2 K(p) 
\nonumber\\ && \hspace{1cm} 
+ m^2\big( 16 m^2-4 p^2-4 q^2 + \frac{p^2 q^2}{m^2}\big) L(p,-q,0)\Big\}
\nonumber\\
-i \Gamma_{\sigma,\pi,\sigma,\pi}^{ab}(q,p,-q) 
&=&\delta_{ab} N\Big\{I(p+q)+I(p-q)+ p^2 (4m^2-q^2) L(p,-q,p-q)\Big\}
\nonumber\\
-i \Gamma_{\sigma,\pi,\pi,\sigma}^{ab}(q,p,-p) 
&=&\delta_{ab} N\Big\{-I(p+q)-I(0)- (4m^2-q^2) (K(q)-p^2 L(p,-q,0))
\nonumber\\ && \hspace{1cm}
+p^2 K(p)-2 p\!\cdot\!q\  M(p,-q)\Big\}
\nonumber\\
-i\Gamma_{\pi,\pi,\pi,\pi}^{abcd}(q,p,-q) 
&=& -N\kappa_{abcd}\Big\{I(p+q)+I(p-q)-p^2q^2 L(p,-q,p-q)\Big\}
\nonumber\\
-i\Gamma_{\pi,\pi,\pi,\pi}^{abcd}(q,p,-p) 
&=& -N\kappa_{abcd}\Big\{I(p+q)+I(0)-p^2 K(p)
\nonumber\\ && \hspace{1.7cm}
-q^2 K(q)+2 p\!\cdot\!q M(p,-q)+p^2 q^2 L(p,-q,0)\Big\}
\nonumber\\
-i\Gamma_{\rho,\sigma,\rho,\sigma}^{ab}(q,p,-q) 
&=& -2\delta_{ab}N\Big\{I(p+q)+I(p-q)+2 I(q) -p\!\cdot\!q (M(p,-q)-M(p,q))
\nonumber\\ && \hspace{1.5cm} 
+(4 m^2-2 p^2)(M(p,q)+M(p,-q))
\nonumber\\ && \hspace{1.5cm}
+m^2 (8 m^2-6 p^2+4 q^2+\frac{p^4-(p\!\cdot\!q)^2}{m^2}) L(p,-q,p-q)\Big\}
\nonumber\\
-i\Gamma_{\rho,\sigma,\sigma,\rho}^{ab}(q,p,-q) 
&=& -2\delta_{ab}N\Big\{-I(p+q)-I(0)+(p^2-4 m^2) K(p)
\nonumber\\ &&\hspace{1.5cm}
+(q^2+2 m^2) K(q) +(4 m^2-2 p\!\cdot\!q) M(p,-q)
\nonumber\\ &&\hspace{1.5cm}
+m^2 (8 m^2-2 p^2+4 q^2-\frac{p^2 q^2}{m^2}) L(p,-q,0)\Big\}
\nonumber\\ 
-i\Gamma_{\rho,\pi,\rho,\pi}^{abcd}(q,p,-q) 
&=& 2N\kappa_{abcd}\Big\{-I(p+q)-I(p-q)-2 I(q)
\nonumber\\ &&\hspace{1.5cm}
+2 p^2 (M(p,q)+M(p,-q)) +p\!\cdot\!q(M(p,-q)-M(p,q)) 
\nonumber\\ &&\hspace{1.5cm}
+(2 m^2 p^2- p^4+(p\!\cdot\!q)^2) L(p,-q,p-q)\Big\}
\nonumber\\
-i\Gamma_{\rho,\pi,\pi,\rho}^{abcd}(q,p,-q) 
&=& 2 N\kappa_{abcd}\Big\{I(p+q)+I(0)-p^2 K(p)-(q^2+2 m^2) K(q)
\nonumber\\ &&\hspace{1.5cm}
+2p\!\cdot\!q M(p,-q)+p^2 (2 m^2+q^2) L(p,-q,0)\Big\}~,
\eea
with $\Gamma_{\rho,M,M,\rho}(q,p,-q) =
g_{\mu\nu}\Gamma_{\rho,M,M,\rho}^{\mu\nu}(q,p,-q),
\Gamma_{\rho,M,\rho,M}(q,p,-p) = g_{\mu\nu}\Gamma_{\rho,M,\rho,M}^{\mu\nu}(q,p,-p)$ and $\kappa_{abcd} = \delta_{ab}\delta_{cd}+\delta_{ad}\delta_{bc}-\delta_{ac}\delta_{bd}$.


\section{Expressions at non-zero temperature}
\label{aptemp}
\setcounter{equation}{0}

To determine the temperature dependence of various quantities we need for the
calculation of the quark condensate at non-zero temperature in \Sec{qqatt} we
adopt the imaginary time or Matsubara formalism (see
e.g. Ref.~\cite{fetter}). In principle this amounts to replace the integration
over energy in the zero temperature expressions by a sum over fermionic
or bosonic Matsubara
frequencies $\omega_n$:
\beq
i \intk f(k)\rightarrow -T\sum_n\int\frac{d^3 k}{(2\pi)^3}
f(i\omega_n,\vec{k})~.
\label{rpt}
\eeq
With this
replacement prescription we can define the temperature analogue to
the elementary integrals, e.g.
\bea
I(p) &=& \intk\frac{1}{(k^2-m^2+i \eps)( (k+p)^2-m^2+
  i\eps)}\rightarrow\nonumber\\
I(i\omega_l,\vec{p})&=&
iT\sum_n\int\frac{d^3k}{(2\pi)^3} \frac{1}
{( (i\omega_n)^2-\vec{k}^2-m^2)
  ( (i\omega_n+i\omega_l)^2-(\vec{k}+\vec{p})^2-m^2)}~.
\eea 

This example also illustrates our notation: At non-zero temperature
the integral depends on energy and  three-momentum separately, which is 
indicated via a second argument. In that way it can be clearly distinguished
from its vacuum counterpart with only one argument.
A similar notation is used for other momentum dependent integrals. 
The non-zero-temperature analogue to the integral $I_1$ will be denoted by 
$I_{1T}$. 

We will now summarize the explicit expressions for various temperature
dependent  quantities which are related to the determination of the quark
condensate at non-zero temperature. 
The temperature analogue to the gap equation \eq{gapex} is given by
\bea
m_T &=& m_0 - 2 g_s 4 N_c N_f m_T T\sum_n\int\frac{d^3 k}{(2 \pi)^3}
 \frac{1}{(i\omega_n)^2-E^2}\nonumber\\
 &=& m_0 + 2 g_s 4 N_c N_f m_T I_{1T}~,
\label{gapt}
\eea  
with $E=\sqrt{\vec{k}^2+m_T^2}$ and $\omega_n = (2n+ 1)\pi T $ being
fermionic Matsubara frequencies. 

The polarization functions for the RPA mesons read 
\bea 
\Pi_{\sigma}(i \omega_l,\vec{p})& =& 4i N_c N_f I_{1T} - 2i N_c N_f
((i\omega_l)^2-\vec{p}^2-4 m_T^2) I(i \omega_l,\vec{p})\nonumber \\
\Pi_{\pi}(i\omega_l,\vec{p})& =& 4i N_c N_f I_{1T} - 2i N_c N_f
((i\omega_l)^2-\vec{p}^2) I(i \omega_l,\vec{p})~,
\eea
with $\omega_l = 2 l \pi T$ being bosonic Matsubara
frequencies. Below the phase transition the integral $I_{1T}$ can
again be replaced with the help of the gap equation \eq{gapt}
(cf. \eqs{sigma} to (\ref{pseudopi})).

Finally, the constant $\Delta_T$ is given by 
\bea
\Delta_T 
&=& 4i N_c N_f\ m_T T \int\frac{d^3p}{(2\pi)^3}\sum_l{\Big\{}
\nonumber\\ &&\hspace{2.3cm}
D_\sigma(i\omega_l,\vec{p})(2I(i\omega_l,\vec{p})+I(0,0)-((i\omega_l)^2-\vec{p^2}-4
m_T^2) K(i\omega_l,\vec{p})) \nonumber \\
&&\hspace{2.cm} +  D_{\pi}(i\omega_l,\vec{p})\  (\;3I(0,0)\;-\;3 ((i\omega_l)^2-\vec{p}^2)\ K(i\omega_l,\vec{p})\;){\Big\}}~, 
\label{delta}
\eea
where $\omega_l$ are again bosonic Matsubara frequencies.
\end{appendix}

\end{document}